\documentclass[12pt]{article}
\usepackage{amsmath, amssymb, amsfonts, graphicx, hyperref, makecell, geometry, subcaption}
\usepackage[
  backend=biber,
  sorting=none,
  style=numeric-comp]{biblatex}
\usepackage{xcolor} 
\newcommand*\update[1]{\textcolor{black}{#1}}
\hypersetup{hidelinks}
\title{Modelling the photocatalytic oxidation of methane and other air pollutants for applications in ventilation systems}
\author{Samuel D. Tomlinson$^\text{1,\,2,\,3}$, Aliki Marina Tsopelakou$^\text{1,\,2}$, Tzia Ming Onn$^\text{1,\,2}$, \\ Steven R. H. Barrett$^\text{1}$, Adam M. Boies$^\text{1,\,2,\,4}$ and Shaun Fitzgerald$^\text{1,\,2}$\\[2ex]}
\date{
    $^\text{1}$Department of Engineering, University of Cambridge, Cambridge, UK\\%
    $^\text{2}$Centre for Climate Repair, Cambridge, UK\\%
    $^\text{3}$School of Computing and Mathematical Sciences, University of Greenwich, London, UK\\%
    $^\text{4}$Department of Mechanical Engineering, Stanford University, Stanford, USA\\[2ex]}

\begin{document}
\maketitle

\begin{abstract}
Photocatalytic oxidation (PCO) is a promising strategy for indoor air purification and outdoor pollutant abatement, potentially offering treatment for climate- and health-relevant pollutants such as methane (CH$_4$), nitrogen oxides (NO$_\text{x}$) and volatile organic compounds (VOCs). 
In this work, we present experiments evaluating the PCO of CH$_4$ (2 to 10 ppm) under varying UV-C light intensities (4 to 59 W/m$^2$), using titanium dioxide (TiO$_2$) as the photocatalyst. 
At 2 ppm CH$_4$, TiO$_2$ achieves a maximum conversion efficiency of 24.4\% and a maximum apparent quantum yield of $0.013$\% over the tested UV-C light intensities, demonstrating activity at environmentally relevant concentrations. 
We develop a model to interpret the experimental results and assess the potential of PCO for ventilation applications. 
The model is validated against our CH$_4$ data and literature results for formaldehyde (HCHO) and NO$_\text{x}$. 
While laboratory-scale configurations achieve high conversions (e.g., 24.4\% for CH$_4$), ventilation-scale performance is predicted to be limited by thin concentration boundary layers and short residence times, with conversion efficiencies dropping to around $0.017$\%. 
Finally, we estimate the climate impact of CH$_4$ removal in terms of CO$_2$e emission rates, demonstrating that TiO$_2$-based PCO in ventilation applications could yield a net climate benefit (i.e., a net-negative CO$_2$e emission rate) when the modelled CO$_2$e removal rate exceeds emissions from catalyst material production and UV-C operation, which could be facilitated by leveraging pre-existing UV-C irradiation. 
\end{abstract}

\section{Introduction}

Air pollution is among the leading environmental and health risks worldwide, contributing significantly to respiratory disease, cardiovascular morbidity and premature mortality~\cite{manisalidis2020environmental, world2021global, sosa2017human}. 
While outdoor air pollution has historically drawn more attention, due in part to its connection with climate change~\cite{kumar2021climate, jones2023national}, recent studies highlight that indoor air quality can be equally hazardous, especially as people now spend over 90\% of their time indoors, in homes, workplaces, schools and transport systems~\cite{malayeri2019modeling}. 
This shift places indoor air quality at the forefront of public health and building design. 
Indoor environments can be polluted with volatile organic compounds (VOCs), nitrogen oxides (NO$_\text{x}$), ozone (O$_3$), methane (CH$_4$) and nitrous oxide (N$_2$O), originating from building materials, cleaning products, combustion sources and infiltration from outdoor air~\cite{manisalidis2020environmental, world2021global}. 
Some of these pollutants are harmful at parts-per-billion levels and have been linked to long-term health effects, including asthma, neurotoxicity and cancer. 
Outdoor air, meanwhile, contains mixtures of anthropogenic and biogenic species. 
CH$_4$ and N$_2$O are potent greenhouse gases (GHGs) with global warming potentials (GWPs) many times greater than carbon dioxide (CO$_2$)~\cite{kumar2021climate, prentice2001carbon}. 
NO$_\text{x}$ and VOCs serve as precursors to secondary pollutants, including tropospheric O$_3$ and fine particulate matter (PM)~\cite{boningari2016impact, zhang2019ozone}. 
Despite regulatory efforts, progress in reducing emissions of these pollutants from sources (e.g., vehicles, industry, agriculture) has been slow. 
To address these diffuse emissions, there is interest in integrating removal technologies (e.g., catalysts) into built environments that experience substantial airflow (e.g., building exteriors, ventilation systems), as part of a distributed approach to air-quality management~\cite{randall2024cost, tomlinson2025harnessing, ma2025}. 

Photocatalytic oxidation (PCO) has emerged as a potentially low-energy and scalable strategy to degrade air pollutants~\cite{zhang2024multivariate, timmerhuis2021connecting}. 
In these systems, a photocatalyst, typically titanium dioxide (TiO$_2$), is irradiated with UV or visible light, generating reactive oxygen species such as hydroxyl radicals (OH$\cdot$) and superoxide anions (O$_2^-$)~\cite{malayeri2019modeling}. 
These reactive oxygen species can oxidise adsorbed pollutants into less harmful or less radiatively active products, without requiring chemical reagents or frequent replacement. 
PCO technologies have been explored in diverse applications, including building exteriors and indoor air purifiers~\cite{ma2025, yu2017experiments, Russell2021Review, ballari2010modelling}. 
Their promise lies in continuous pollutant degradation under natural or artificial illumination. 
However, widespread adoption remains limited by uncertainties in performance under realistic conditions, where pollutant levels are low, airflows are complex and operational constraints such as humidity and light intensity vary~\cite{timmerhuis2021connecting, yu2017experiments}.
Beyond catalyst material selection and operating conditions, the overall feasibility of pollutant-removal technologies also depends on fabrication, installation and energy inputs, all of which influence the net climate impact of deployed systems.

The performance of PCO has been shown to be sensitive to operating conditions~\cite{malayeri2019modeling, zhang2024multivariate, timmerhuis2021connecting, yu2017experiments, ballari2010modelling}, including air pollutant concentration, light intensity, humidity, catalyst properties and flow rate. 
Bridging the gap between laboratory-scale performance and effectiveness in application will require models that incorporate both reaction kinetics and mass transport and are validated against multiple experimental measurements for different air pollutants. 
It is critical to distinguish between reaction-limited and mass-transfer-limited regimes~\cite{zhang2024multivariate, timmerhuis2021connecting}, which can exhibit similar behaviour but require different design strategies (e.g., improving catalyst activity in the former, or enhancing flow conditions in the latter). 
Overly simplified models may obscure these distinctions, whereas over-parametrised models or complicated numerical simulations can lack physical interpretability and may not readily support optimisation. 

Recent theoretical and experimental studies have begun to address some of these issues~\cite{malayeri2019modeling, zhang2024multivariate, timmerhuis2021connecting, yu2017experiments, ballari2010modelling}. 
Zhang et al.~\cite{zhang2024multivariate} developed a model coupling the Navier--Stokes equations, advection--diffusion transport and a Langmuir--Hinshelwood kinetic model, to predict NO$_\text{x}$ removal using UiO-66-NH$_2$ across varying light intensities, pollutant concentrations and humidity levels. 
Timmerhuis et al.~\cite{timmerhuis2021connecting} investigated the role of the P\'{e}clet and Damk\"{o}hler numbers in characterising Bisphenol A removal via TiO$_2$, demonstrating that neglecting mass transport can lead to errors in rate constant predictions. 
Malayeri et al.~\cite{malayeri2019modeling} reviewed reactor configurations and photocatalysts for VOC removal, highlighting the role of catalyst exposure, photon-flux distribution and boundary-layer effects in system performance. 
Yu et al.~\cite{yu2017experiments} investigated a Trombe-wall PCO system for formaldehyde (HCHO) removal using a TiO$_2$-coated ceramic and developed a kinetic model accounting for UV intensity, temperature, humidity and concentration. 
Ballari et al.~\cite{ballari2010modelling} combined experimental and theoretical studies of NO$_\text{x}$ removal using TiO$_2$-coated concrete pavers and used a Langmuir--Hinshelwood kinetic model to capture the effects of irradiance, humidity and flow conditions. 
Collectively, these studies highlight the variety of operational conditions and the effectiveness of PCO systems at laboratory scale. 

Russell et al.~\cite{Russell2021Review} conclude that deployments of TiO$_2$ coatings in outdoor applications typically yield ambient pollutant concentration reductions of around 2\% in the immediate vicinity of treated surfaces. 
In addition, formation of secondary species has been observed in some studies, such that air-quality impacts depend on surface reactivity and atmospheric chemistry.
Cost analyses suggest that using photocatalytic oxidation for atmospheric CH$_4$ removal is prohibitively expensive due to the high flow rates required for effective removal~\cite{randall2024cost, Pennacchio2024, hickey2024economics}.
Further analysis shows that targeting higher-concentration CH$_4$ sources can yield meaningful climate benefits at source concentrations above 10 ppm, with cost-effective operation using current technologies requiring concentrations above $10^{3}$ ppm~\cite{abernethy2023assessing}. 
Chlorine and OH$\cdot$ radicals can oxidise CH$_4$ in the gas phase, and laboratory photochemical systems have reported relatively low energy inputs for low-concentration pollutant removal~\cite{randall2024cost, krogsboll2025efficient}. 
Photochemical destruction approaches have also been investigated for N$_2$O and fluorinated anaesthetic gases in exhaust streams~\cite{rauchenwald2020new}.

Building on these foundations, the present study combines modelling and experiments to assess the performance of PCO systems for ventilation applications. 
We conduct CH$_4$-oxidation experiments under varying UV-C light intensities at environmentally relevant CH$_4$ concentrations for indoor and outdoor air.  
These measurements inform the development of a model, solved using asymptotic and numerical methods, that couples non-linear Langmuir--Hinshelwood kinetics with advection--diffusion transport under UV-C illumination. 
We validate the model using our CH$_4$ experimental data, as well as literature data for NO$_\text{x}$ and HCHO oxidation using TiO$_2$, over a range of flow rates, concentrations and light intensities. 
The model allows us to quantify the extent to which laboratory- and application-scale PCO systems are constrained by concentrations, boundary-layer thickness, residence time and geometry. 
We estimate the potential removal rates and the associated CO$_2$e emissions rates for the PCO of CH$_4$, enabling assessment of PCO performance under realistic operating conditions and providing a framework for the design and optimisation of potential air-purification technologies. 

The rest of the paper is organised as follows. 
In Section~\ref{sec:Experiments}, we describe the experimental setup and measurements for CH$_4$. 
Section~\ref{sec:Model} presents the model, discusses the numerical methods and parameter fitting. 
In Section~\ref{sec:Results}, we present model and experimental results, investigating performance at lab and application scales, and estimating the potential climate benefits of PCO technologies for CH$_4$ removal. 
Section~\ref{sec:Discussion} summarises the results and applications of this modelling and measurement framework, and discusses potential future extensions. 

\section{Experiments} \label{sec:Experiments}

\begin{figure}[t!]  
\centering
    \includegraphics[width=\linewidth]{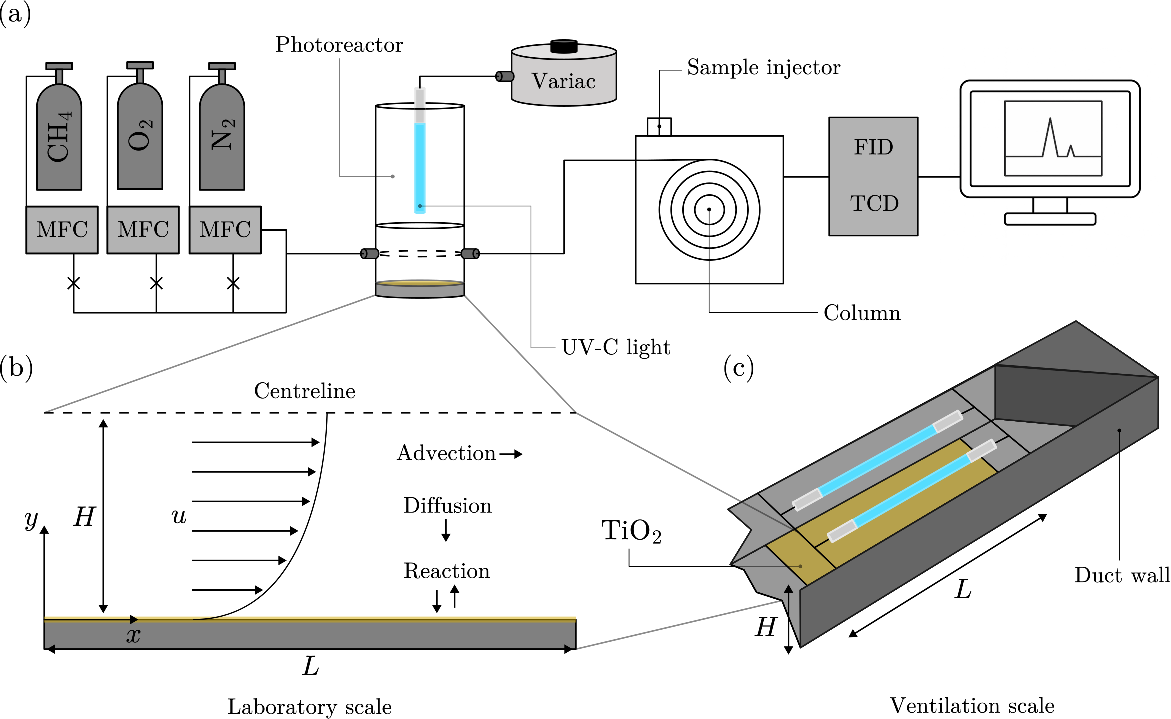}
    \caption{(a) Schematic of the laboratory setup used for the PCO experiments, showing the gas delivery system, UV-C illumination, reactor, GC, FID and TCD for CH$_4$ removal over TiO$_2$. 
    (b) Two-dimensional reactor model, representing laminar flow with advection, diffusion and reaction at the UV-C illuminated photocatalyst wall. 
    (c) A potential ventilation application of this UV-C pollutant-removal technology and its connection to our laboratory setup.}
    \label{fig:composite}
\end{figure}

To evaluate the PCO of CH$_4$ at environmentally relevant concentrations (typical of indoor and outdoor air), we designed a set of experiments informed by previous studies on catalyst performance~\cite{tsopelakou2024exploring, marina2026evaluating}. 
Comparative data for NO$_\text{x}$ and HCHO are taken from Ballari et al.~\cite{ballari2010modelling} and Yu et al.~\cite{yu2017experiments}, respectively, and are discussed further in Section~\ref{sec:Pollutant removal under variable conditions}. 
Experimental conditions, PCO measurements and the flow--reactor system are described in Section~\ref{sec:Experimental Conditions}. 
Figure~\ref{fig:composite}(a) shows the experimental setup used to measure CH$_4$ conversion efficiency. 
\update{The kinetic parameters fitted to each dataset are described in Section~\ref{sec:Model} and their values are listed in Table~\ref{ref:parameters_1}.}

\subsection{Experimental setup and conditions} \label{sec:Experimental Conditions}
\vspace{.2cm}

To evaluate the PCO of CH$_4$ at low concentrations relevant to air-purification applications, titanium dioxide (TiO$_2$, Titanium (IV) oxide, anatase, nanopowder, \textless 25 nm, 99.7\%, trace metals basis, Sigma-Aldrich) was selected as a representative photocatalyst. 
TiO$_2$ is widely studied for its strong photocatalytic activity under UV-C illumination~\cite{malayeri2019modeling, tsopelakou2024exploring}, making it a suitable benchmark for evaluating CH$_4$ oxidation rates. 
In addition to PCO performance, the broader climate benefit of PCO systems in potential ventilation applications is quantified by considering the energy consumption of the UV-C illumination and the catalyst’s embodied carbon, as discussed in Section~\ref{sec:Climate benefit for UV light} and detailed in Appendix~\ref{sec:Methodology for photocatalyst}. 
Practical deployment of PCO systems would be most effective when high photocatalytic activity is combined with low overall carbon intensity. 

The PCO experiments were performed using a reactor system designed to evaluate CH$_4$ removal at environmentally relevant concentrations under UV-C irradiation (Figure~\ref{fig:composite}a). 
For each experiment, 2 g of TiO$_2$ was deposited as a thin layer covering the bottom surface of the cylindrical reactor (38 cm total height, 6 cm diameter) and irradiated by a UV-C lamp delivering light intensities between 4 and 59 W/m$^2$ on the catalyst surface. 
The 2D geometry adopted in the model (Section~\ref{sec:Model}) is an idealisation of this configuration, with the centreline height (1 cm) chosen to preserve the residence time of the experimental setup.
CH$_4$ and O$_2$ served as the active reactants, with N$_2$ the carrier gas. 
Gas flows were regulated using mass flow controllers (MFCs), with the total volumetric flow rate maintained at 100 mL/min for all experiments. 
This value falls within the range of flow rates reported for dilute methane abatement systems in recent literature~\cite{Pennacchio2024}.
Input CH$_4$ concentrations ranged from 2 to 10 ppm, reflecting environmentally relevant levels for indoor and outdoor air. 
Experiments were conducted at ambient temperature (20\,\textdegree{}C) under dry conditions (no added humidity). 
\update{UV-C illumination was selected because the narrow spectral output of the lamp simplifies photon-flux calibration and AQY calculations, and because UV-C germicidal lamps represent the most widely deployed UV infrastructure in existing HVAC systems, making the results relevant to the deployment scenario described in Section~\ref{sec:Full-duct simulation and scale-up}. 
TiO$_2$ is also active under UV-A illumination; the model structure is expected to carry over, but extrapolation to UV-A conditions would require separate calibration of model parameters, since TiO$_2$'s photon absorption efficiency is wavelength-dependent.} 

\subsection{Measurement and performance metrics}

Gas-phase products were quantified using an Agilent 8890 gas chromatograph (GC) equipped with a flame ionisation detector (FID) and a thermal conductivity detector (TCD). 
Peak areas for CH$_4$ and CO$_2$ were extracted from the chromatograms and converted to concentrations in parts per million (ppm). 
The CH$_4$ conversion efficiency was calculated assuming complete oxidation to CO$_2$. 
This assumption is supported by the absence of additional FID peaks in our chromatograms. 
The conversion efficiency is given by
\begin{equation} 
\eta = \frac{C(t=0) - C(t=T)}{C(t=0)}, 
\label{eq:conversion}
\end{equation}
\update{where $C(t=0)$ and $C(t=T)$ are the CH$_4$ concentrations measured at the outlet before and during UV-C irradiation, respectively.} 
\update{In the absence of UV-C irradiation, no catalytic reaction occurs and the CH$_4$ concentration is uniform throughout the reactor at steady state, so $C(t=0)$ serves as an experimental proxy for the inlet concentration $c_\text{in}$, making $\eta$ directly comparable to the model prediction $\eta^{\text{model}}$ in~\eqref{eq:removal_eff}.} 
This stoichiometric approach provides a measure of CH$_4$-degradation efficiency under the tested CH$_4$ concentrations and light intensities. 
The apparent quantum yield (AQY) is defined as the ratio of the CH$_4$ reaction rate to the incident UV-C photon flux (both in moles/(m$^2$s)) and is computed for each combination of CH$_4$ concentration and light intensity as 
\begin{equation} 
\text{AQY} = \frac{r_{\text{CH$_4$}} }{r_\text{{Photon}}}.
\label{eq:AQY}
\end{equation}
In the above, $r_\text{CH$_4$}$ is determined from the measured CH$_4$ conversion efficiency at each CH$_4$ concentration and $r_\text{{Photon}} = I /(E_\text{{Photon}} N_A) $  is the incident UV-C photon flux (moles/(m$^2$s)), where $I$ is the incident light intensity (W/m$^2$), $E_\text{{Photon}}$ is the energy of photons (J) and $N_{A}$ is Avogadro's constant (mol$^{-1}$).

In addition to the surface reaction at the TiO$_2$ catalyst, UV-C irradiation at the wavelengths emitted by our UV-C lamp can photolyse O$_2$ to generate reactive oxygen species, including OH$\cdot$, which are capable of oxidising CH$_4$ in the bulk~\cite{krogsboll2025efficient, Pennacchio2024, iversen2025mitigation}.
To isolate the catalytic contribution from the bulk reaction and enable comparison with the reactor PCO model (Section~\ref{sec:Model}), a control experiment was performed for each combination of CH$_4$ concentration and UV-C light intensity, repeating the measurement with no catalyst present.
The conversion efficiency is then defined as
\begin{equation}
\eta^{\text{exp}} = \eta^{\text{cat}} - \eta^{\text{no-cat}},
\label{eq:catalytic_conversion}
\end{equation}
where $\eta^{\text{cat}}$ is the total conversion efficiency measured with catalyst present 
and $\eta^{\text{no-cat}}$ is the conversion efficiency measured without catalyst present, both evaluated using~\eqref{eq:conversion}.
We use $\eta^{\text{exp}}$ to characterise catalyst performance and to compute the AQY via~\eqref{eq:AQY}; these quantities are compared with reactor-model predictions in Section~\ref{sec:Pollutant removal under variable conditions}.
A detailed quantification of the total conversion efficiency, including the bulk and surface reactions, as well as a description of the bulk-phase reaction mechanism, is left for future work.

\subsection{PCO mechanism}

\update{The PCO mechanism on TiO$_2$ proceeds through band-gap excitation, generating electron--hole pairs.
Photogenerated holes react with surface-adsorbed water and hydroxyl groups to form hydroxyl radicals (OH$\cdot$), while electrons react with O$_2$ to form superoxide anions (O$_2^{-}\cdot$)~\cite{malayeri2019modeling}.
These reactive oxygen species subsequently attack adsorbed CH$_4$ molecules.
For CH$_4$, the initial C--H bond activation by OH$\cdot$ is rate-limiting owing to the high bond dissociation energy; successive oxidation steps produce methanol (CH$_3$OH), formaldehyde (HCHO), formic acid (HCOOH) and ultimately CO$_2$ and H$_2$O~\cite{pitchai1986}.}

\begin{table}[t!]
\centering
\caption{Summary of parameters for each of the PCO studies on CH$_4$, HCHO and NO$_\text{x}$ removal using TiO$_2$. 
Quantities marked with an asterisk are fitted to the experimental data. 
Parameters held fixed across all studies are: catalyst (TiO$_2$), molecular diffusivity ($1.8 \times 10^{-5}$ m$^2$/s), dynamic viscosity ($1.85 \times 10^{-5}$ Pa$\cdot$s) and pressure (1 atm). 
}
\vspace{.25cm}
\renewcommand{\arraystretch}{1.2}
\small
\begin{tabular}{|l|c|c|c|}
\hline
& Ballari et al.~\cite{ballari2010modelling} & Yu et al.~\cite{yu2017experiments} & Present work \\
\hline
\multicolumn{4}{|c|}{Flow and geometric quantities} \\
\hline
Pollutant & NO$_\text{x}$ & HCHO & CH$_4$ \\
Concentration (ppm) & 0.11 to 1.09 & 0.17 to 0.9 & 2 to 10 \\
Centreline height (mm) & 1.5 & 5 & 10 \\
Length (mm) & 200 & 240 & 60 \\
Flow rate (L/min) & 3 to 5 & 1 to 5 & 0.1 \\
Temperature (\,\textdegree{}C) & 20 & 15 to 45 & 20 \\
Humidity (\%) & 10 to 80 & 20 to 90 & 0 \\
UV-C (W/m$^2$) & 0.3 to 13 & 2.1 to 9 & 4 to 59 \\
\hline
\multicolumn{4}{|c|}{Fitted constants} \\
\hline
Light$^*$ & 0.23 & 0.38 & 0.21 \\
Adsorption$^*$ (m$^3$/mol) & $3.9\times10^{-5}$ & $1.1\times10^{-5}$  & $4.6\times10^{-5}$ \\
Rate$^*$ (mol$\cdot$m$^{2(a-1)}$/(s$\cdot$W$^{a}$)) & 5.22 & 1.48 & 0.084 \\
\hline
\multicolumn{4}{|c|}{Non-dimensional numbers} \\
\hline
P\'{e}clet & 2.50 to 4.17 & 2.50 & 0.085 \\
Damk\"{o}hler & 3.45 & 0.26 to 0.46 & 0.016 to 0.027 \\
Langmuir & 0.54 to 5.71 & 0.26 to 1.36 & 0.17 to 0.84 \\
\hline
\end{tabular}
\label{ref:parameters_1}
\end{table}

\subsection{Experimental scope and limitations}

A schematic of the PCO setup, including the gas delivery system, UV-C lamp alignment, reactor chamber, GC, FID and TCD analysis configuration, is shown in Figure~\ref{fig:composite}(a). 
Table~\ref{ref:parameters_1} summarises the parameters for the CH$_4$ experiments, as well as for HCHO~\cite{yu2017experiments} and NO$_\text{x}$~\cite{ballari2010modelling} PCO systems. 
All CH$_4$ experiments were conducted under controlled laboratory conditions using dry gas feeds, clean reactant streams and an idealised reactor geometry. 
While these conditions allow for reproducible catalyst performance, they do not fully replicate real-world environments, where humidity, contaminants, fluctuating flow and light-intensity profiles may affect the catalyst oxidation state and long-term stability~\cite{Anekwe2025, Dissanayake1991}. 
Accordingly, the reported performance metrics reflect catalytic behaviour under laboratory conditions rather than application environments.

\update{The dry experimental conditions used in the present CH$_4$ study do not capture the humidity dependence reported for photocatalytic CH$_4$ oxidation, where competition for adsorption sites and enhanced OH$\cdot$ generation produce a non-monotonic effect on conversion efficiency~\cite{ying2025unraveling}; by contrast, the NO$_\text{x}$ and HCHO datasets used for validation here~\cite{yu2017experiments, ballari2010modelling} report a monotonic decrease in conversion efficiency with increasing relative humidity. 
These effects will be incorporated in future studies at representative indoor humidity levels.}
\update{Catalyst mass and coated length were held fixed across all CH$_4$ experiments; the effect of catalyst area is captured through the reactor length $L$ and fitted parameters $k'$ and $K$ in~\eqref{eq:LH}.
Conversion efficiency increases with $L$ at fixed channel height and flow rate, and is investigated in the reactor-to-duct scaling in Section~\ref{sec:Flow field and concentration distribution}--\ref{sec:Full-duct simulation and scale-up}.}
\update{Residence time effects are captured by the P\'{e}clet number and reactor aspect ratio $L/H$; the experimentally tested flow rate of 100 mL/min corresponds to a low-$\textrm{Pe}$ regime (Table~\ref{ref:parameters_1}), and the ventilation-scale extrapolation in Section~\ref{sec:Full-duct simulation and scale-up} evaluates performance at high-$\textrm{Pe}$ conditions representative of building airflows.}

\section{Model} \label{sec:Model}

With the PCO experiments for CH$_4$ established in Section~\ref{sec:Experiments}, we develop a modelling framework to interpret these results and selected literature data for HCHO and NO$_\text{x}$~\cite{yu2017experiments, ballari2010modelling}. 
The modelled domain is outlined in Figure~\ref{fig:composite}(b). 
In Section~\ref{sec:Full-duct simulation and scale-up}, we scale these results up to ventilation applications to assess potential climate benefits, as illustrated in Figure~\ref{fig:composite}(c). 
We begin by presenting the governing equations and outlining the model assumptions (Section~\ref{sec:Governing equations}--\ref{sec:Non-dimensional problem}). 
Then, we describe the solution approach, which combines asymptotic methods (Appendix \ref{sec:Asymptotic analysis}) and numerical methods (Section~\ref{sec:Numerical methods}). 
Finally, we describe the fitting of model parameters to experimental data (Section~\ref{sec:Experimental systems and kinetic model fitting}). 

\subsection{Governing equations} \label{sec:Governing equations}

We model the transport and removal of a gaseous pollutant in a 2D, steady-state, laminar-flow reactor (Figure~\ref{fig:composite}b). 
The channel has centreline height $H$ and length $L$, with the bottom wall ($y = 0$) coated with a photocatalyst and uniformly illuminated by UV-C light. 
We assume that the flow is fully developed in the streamwise direction and incompressible, with a parabolic streamwise velocity profile and negligible wall-normal velocity. 

Under these assumptions, the pollutant concentration field, $c=c(x, \, y)$, satisfies the steady advection--diffusion equation: 
\begin{equation}
    u \frac{\partial c}{\partial x} = D \frac{\partial^2 c}{\partial y^2},
    \label{eq:advdiff}
\end{equation}
where $u=u(y)$ is the velocity profile and $D$ is the molecular diffusivity. 
The velocity profile for fully-developed laminar flow between parallel plates is
\begin{equation}
    u = \frac{3 \langle u \rangle}{2} \bigg( 1 - \bigg( \frac{y - H}{H} \bigg)^2 \bigg),
    \label{eq:parabolic}
\end{equation}
where $\langle u \rangle$ is the mean streamwise velocity. 
The pollutant enters the reactor at a uniform concentration $c_{\mathrm{in}}$. 
The top wall ($y = 2 H$) is impermeable, while the bottom wall ($y = 0$) undergoes photocatalytic reaction. 
The corresponding boundary conditions are
\begin{equation}
    c(x = 0, \,y) = c_{\mathrm{in}}, \quad
    D \frac{\partial c}{\partial y}(x, \,0) = r(x), \quad
    \frac{\partial c}{\partial y}(x, \,2H) = 0, \label{eq:topwall}
\end{equation}
where we assume, following Zhang et al.~\cite{zhang2024multivariate}, that the surface reaction rate $r=r(x)$ follows a light-modulated Langmuir--Hinshelwood expression:
\begin{equation}
    r(x) = \frac{k' I^a K c(x, \,0)}{1 + K c(x, \,0)},
    \label{eq:LH}
\end{equation}
where $k'$ is the rate constant (mol$\cdot$m$^{2(a-1)}$/(s$\cdot$W$^{a}$)), $K$ is the adsorption equilibrium constant (m$^3$/mol), $I$ is the UV irradiance (W/m$^2$) and $a$ is the light-intensity exponent. 
The conversion efficiency is defined as the reduction in the average pollutant concentration across the channel at the outlet:
\begin{equation}
\eta^{\text{model}} = 1 - \frac{1}{c_{\text{in}}}\int_{y = 0}^{2H} c(x = L, \,y) \, dy. 
\label{eq:removal_eff}
\end{equation}
The unknown parameters in \eqref{eq:LH}, $k'$, $K$ and $a$, are fitted to three experimental datasets, as described in Section~\ref{sec:Experimental systems and kinetic model fitting}. 
The model captures both reaction saturation at high pollutant concentrations and a sub-linear dependence on light intensity, as will be observed in the CH$_4$ experiments (Section~\ref{sec:Experiments}). 

\subsection{Non-dimensionalisation} \label{sec:Non-dimensional problem}

To compare PCO performance across reactors and against results from prior experiments (Section~\ref{sec:Experiments}; Ballari et al.~\cite{ballari2010modelling}; Yu et al.~\cite{yu2017experiments}), we introduce the non-dimensional variables
\begin{equation} \label{eq:non_dim}
x = H \hat{x}, \quad y = H \hat{y}, \quad c = c_{\mathrm{in}} \hat{c}, \quad u = \langle u \rangle \hat{u}. 
\end{equation}
Substituting \eqref{eq:non_dim} into the advection--diffusion equation \eqref{eq:advdiff} yields the transport equation
\begin{equation} \label{eq:non_dim_1}
    \hat{u} \frac{\partial \hat{c}}{\partial \hat{x}} = \frac{1}{\mathrm{Pe}} \frac{\partial^2 \hat{c}}{\partial \hat{y}^2},
\end{equation}
where the P\'{e}clet number, $\mathrm{Pe} = \langle u \rangle H/D$, quantifies the ratio of advective to diffusive transport. 
The velocity field \eqref{eq:parabolic} becomes
\begin{equation}
    \hat{u} = \frac{3}{2} \left( 1 - (\hat{y} - 1)^2 \right). 
    \label{eq:dimless_u}
\end{equation}
The boundary conditions \eqref{eq:topwall} become
\begin{align} 
    \hat{c}(0, \,\hat{y}) = 1, \quad
    \frac{\partial \hat{c}}{\partial \hat{y}}(\hat{x}, \,2) = 0, \quad
    \frac{\partial \hat{c}}{\partial \hat{y}}(\hat{x}, \,0) = \frac{\mathrm{Da} \, \hat{c}(\hat{x}, \,0)}{1 + \beta \hat{c}(\hat{x}, \,0)}, \label{eq:non_dim_2}
\end{align}
where the Damk\"{o}hler number, $\mathrm{Da} = H k' I^a K/D$, measures the ratio of surface reaction rate to mass transport, and the Langmuir parameter, $\beta = K c_{\mathrm{in}}$, quantifies surface saturation at the inlet concentration. 
The conversion efficiency \eqref{eq:removal_eff} is given by
\begin{equation}
\eta^{\text{model}} = 1 - \int_{\hat{y} = 0}^{2} \hat{c}(\hat{x} = 1, \,\hat{y}) \, d\hat{y}. 
\label{eq::non_dim_3}
\end{equation}
Asymptotic solutions to the governing equations and conversion efficiency \eqref{eq:non_dim_1}--\eqref{eq::non_dim_3} are derived in the limits of strong and weak diffusion ($\mathrm{Pe}\ll1$, $\mathrm{Pe}\gg1$), reaction ($\mathrm{Da}\ll1$, $\mathrm{Da}\gg1$) and adsorption ($\beta\ll1$, $\beta\gg1$) in Appendix~\ref{sec:Asymptotic analysis}. 

\subsection{Numerical methods} \label{sec:Numerical methods}

The non-dimensional system of equations, \eqref{eq:non_dim_1}--\eqref{eq:non_dim_2}, is solved numerically using MATLAB’s pdepe~\cite{matlab2025}. 
The $\hat{y}$ domain is discretised using finite differences, and the resulting system is integrated in $\hat{x}$ to obtain the concentration field $\hat{c}$. 
Mesh independence is verified by grid refinement, and standard grid sizes of $N_{\hat{x}} = N_{\hat{y}} = 100$ are used for all results in Section~\ref{sec:Results}. 
The numerical solution is benchmarked against the asymptotic results derived in Appendix~\ref{sec:Asymptotic analysis} and against literature data from Ballari et al.~\cite{ballari2010modelling} and Yu et al.~\cite{yu2017experiments}. 

\subsection{Model fitting to experiments} \label{sec:Experimental systems and kinetic model fitting}

We compare the PCO of three representative indoor and outdoor pollutants: CH$_4$, HCHO and NO$_\text{x}$, across three experimental datasets, each employing TiO$_2$ as the photocatalyst. 
These systems differ in geometry, flow rate, pollutant concentration, operating temperature and UV intensity, as summarised in Table~\ref{ref:parameters_1}. 
\update{The Ballari et al.~\cite{ballari2010modelling} and Yu et al.~\cite{yu2017experiments} datasets are the most appropriate for model validation at the concentration range and geometry considered here.}
Each PCO system is modelled using the 2D advection--diffusion equation for pollutant concentration \eqref{eq:advdiff}, coupled to a surface reaction following light-modulated Langmuir--Hinshelwood kinetics \eqref{eq:topwall}. 
Kinetic parameters ($k'$, $K$, $a$) are estimated by minimizing the sum of squared errors (SSE) between experimental and modelled removal efficiencies across $N$ experimental realisations:
\begin{equation}
\text{SSE}(k', \,K, \,a) = \sum_{i=1}^N \left(\eta^{\text{exp}}_i - \eta^{\text{model}}_i(k', \,K, \,a) \right)^2. 
\end{equation}
The PDE system \eqref{eq:advdiff}--\eqref{eq:topwall} is solved numerically using MATLAB's pdepe, and the kinetic parameters are obtained by minimising the SSE using MATLAB's fmincon~\cite{matlab2025}. 

\section{Results} \label{sec:Results}

\begin{figure}[t!]
\centering
(a) \hfill (b) \hfill (c) \hfill \hfill \hfill \\
\hfill\includegraphics[width=0.33\textwidth,trim={0cm 0 0.9cm 0cm},clip]{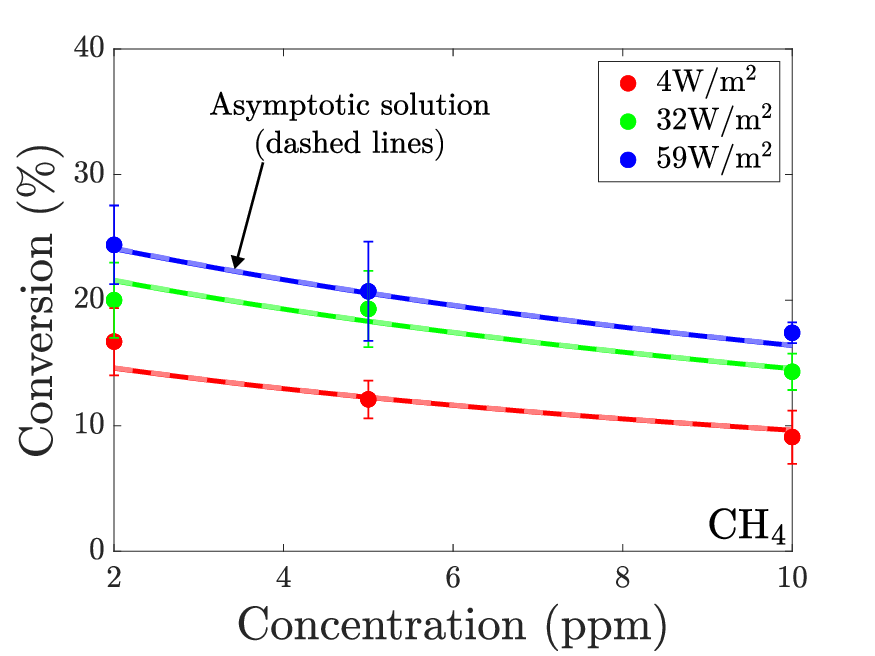}\hfill\includegraphics[width=0.33\textwidth,trim={0 0 0.9cm 0},clip]{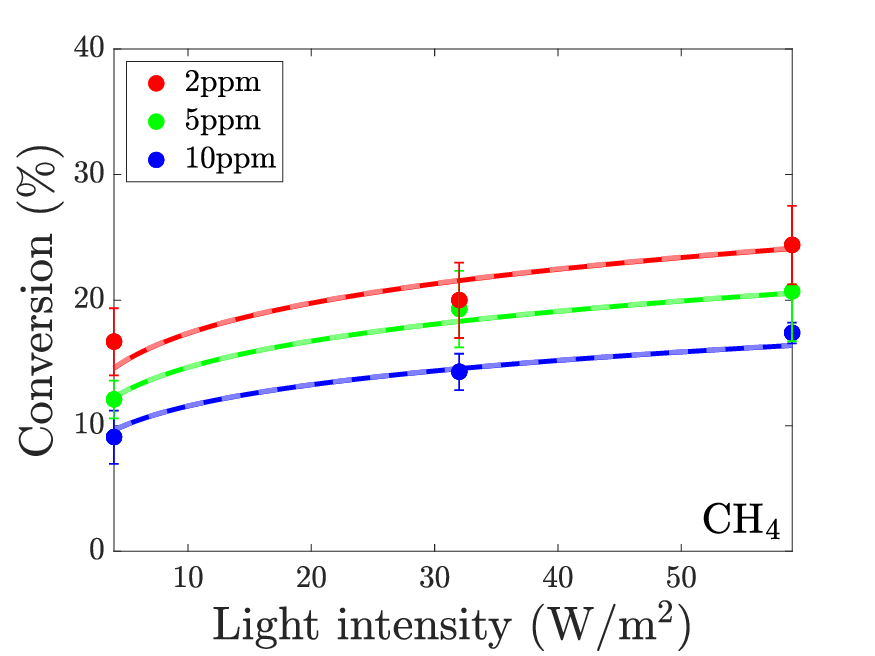}\hfill\includegraphics[width=0.33\textwidth,trim={0 0 0.9cm 0},clip]{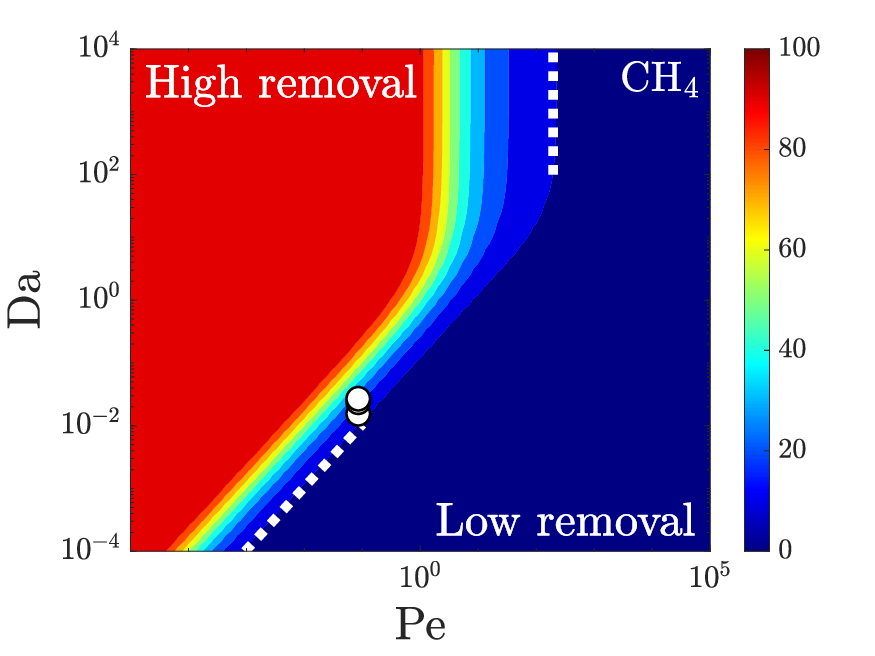} \\
(d) \hfill (e) \hfill (f) \hfill \hfill \hfill \\
\hfill\includegraphics[width=0.33\textwidth,trim={0 0 0.9cm 0},clip]{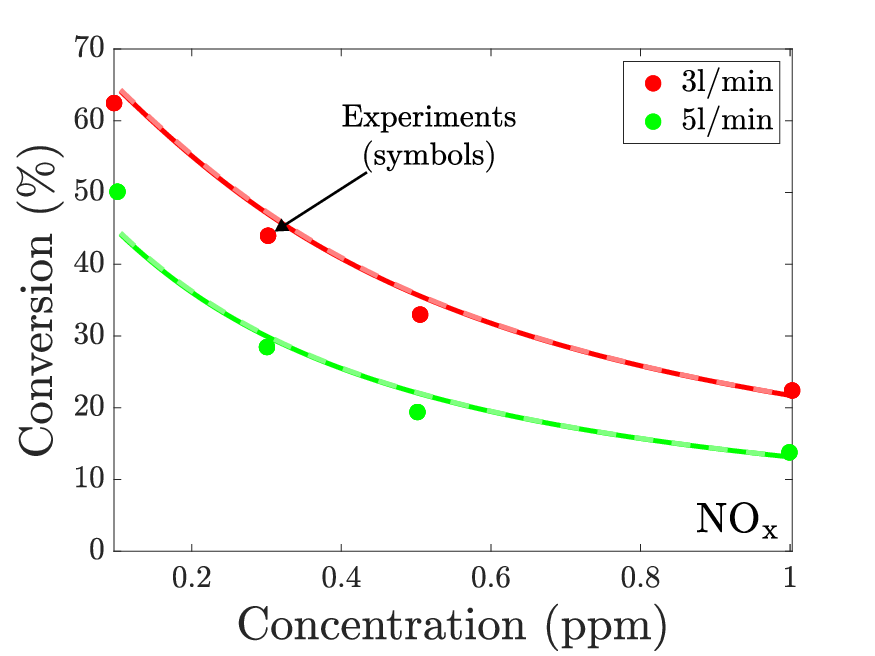}\hfill\includegraphics[width=0.33\textwidth,trim={0 0 0.9cm 0},clip]{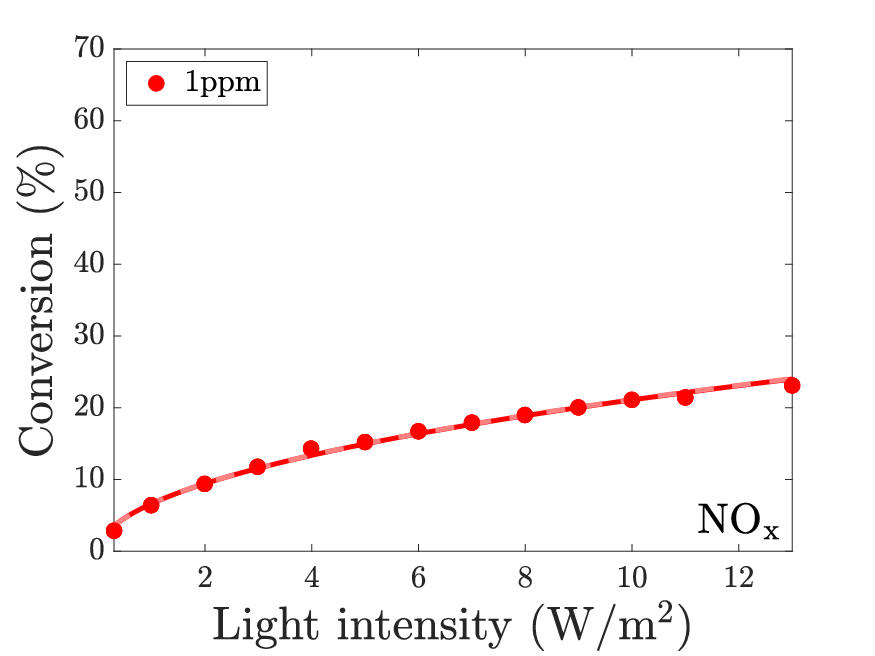}\hfill\includegraphics[width=0.33\textwidth,trim={0 0 0.9cm 0},clip]{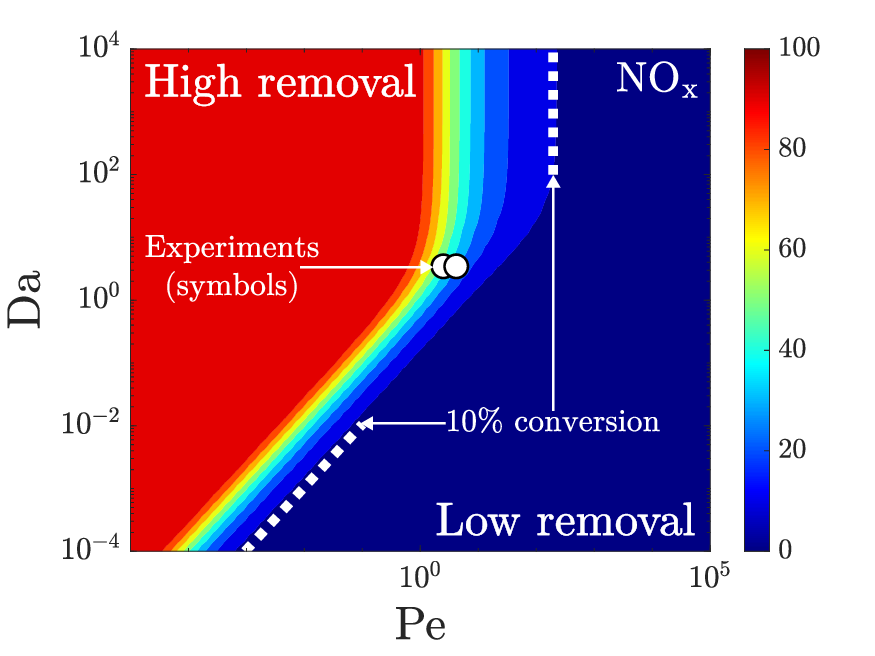} \\
(g) \hfill (h) \hfill (i) \hfill \hfill \hfill \\
\hfill\includegraphics[width=0.33\textwidth,trim={0 0 0.9cm 0},clip]{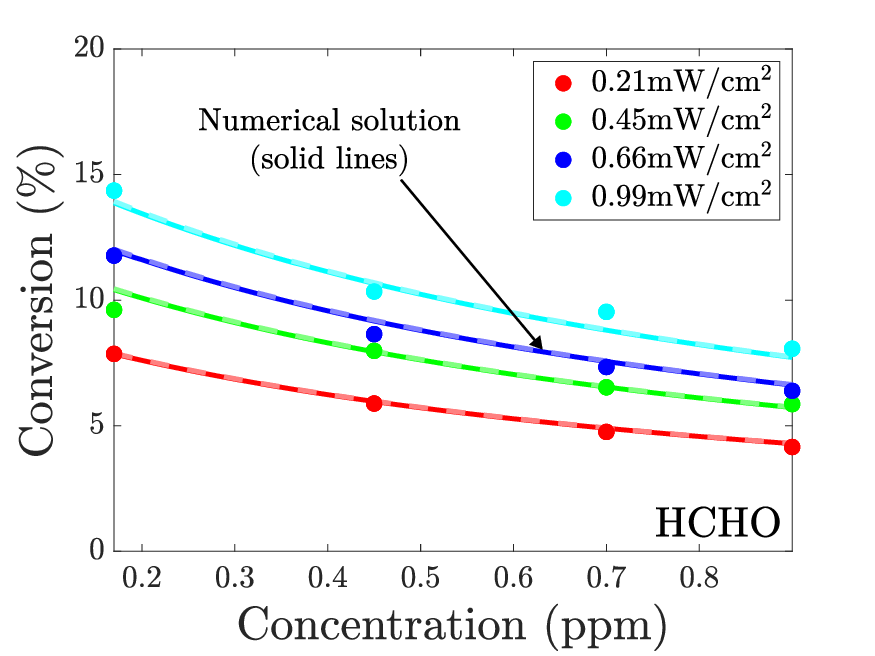}\hfill\includegraphics[width=0.33\textwidth,trim={0 0 0.9cm 0},clip]{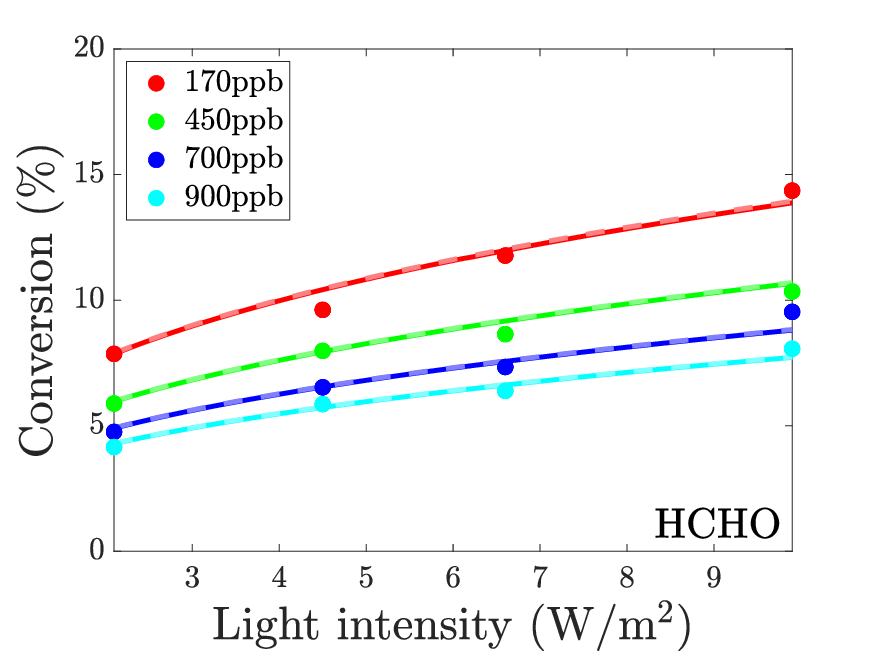}\hfill\includegraphics[width=0.33\textwidth,trim={0 0 0.9cm 0},clip]{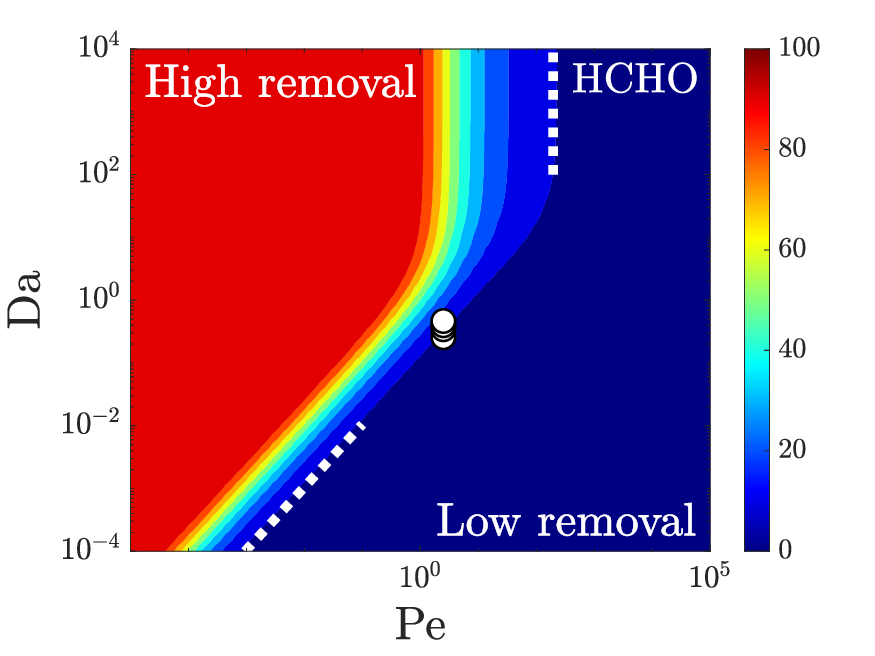} \\
\caption{Conversion efficiency $\eta$ (\%) as a function of (a, d, g) inlet concentration (ppm) and (b, e, h) UV light intensity (W/m$^2$), with the corresponding experimental locations in $(\mathrm{Da}, \mathrm{Pe})$ space and contours of $\eta$ from 0\% (blue) to 100\% (red) shown in (c, f, i). 
Data correspond to (a--c) CH$_4$ (Section~\ref{sec:Experiments}), (d--f) NO$_\text{x}$~\cite{ballari2010modelling}, and (g--i) HCHO~\cite{yu2017experiments}. 
Solid lines denote numerical solutions of the advection--diffusion equation with Langmuir--Hinshelwood kinetics \eqref{eq:advdiff}--\eqref{eq:topwall}, while dashed lines indicate asymptotic solutions in the small-$\textrm{Pe}$ \eqref{eq:rem_eff_small_Pe} and large-$\textrm{Pe}$ limits \eqref{eq:removal_eff_large_Da_final}. 
}
\label{fig:removal_map}
\end{figure}

With the CH$_4$ PCO experiments described in Section~\ref{sec:Experiments} and the model outlined in Section~\ref{sec:Model}, we compare model and experimental trends across CH$_4$, NO$_\text{x}$ and HCHO (Section~\ref{sec:Pollutant removal under variable conditions}), we compare experimental trends for CH$_4$ with others found in the literature (Section~\ref{subsec:Apparent quantum yield}), we scale trends from laboratory reactors to ventilation systems (Section~\ref{sec:Flow field and concentration distribution}--\ref{sec:Full-duct simulation and scale-up}, Figure~\ref{fig:composite}b--c), and we estimate the potential climate benefits of PCO technologies in ventilation systems (Section~\ref{sec:Climate benefit for UV light}). 
All conversion efficiencies, AQYs, and climate-benefit estimates for CH$_4$ reported in this section characterise only the catalytic contribution, isolated from the bulk reaction through the control experiments described in Section~\ref{sec:Experiments}; total values, including the bulk reaction, will be characterised in future work.
\update{Error bars on experimental data points in Figure~\ref{fig:removal_map} indicate the range of conversion efficiencies measured across repeated experiments.}

\subsection{Pollutant removal under variable conditions} \label{sec:Pollutant removal under variable conditions}

Using the PCO model in Section~\ref{sec:Model}, we simulate the conversion efficiency of CH$_4$, NO$_\text{x}$ and HCHO across a range of inlet concentrations (0.11 to 10 ppm) and UV-C light intensities (0.1 to 59 W/m$^2$), consistent with the experimental conditions (Section~\ref{sec:Experiments}, \cite{ballari2010modelling}, \cite{yu2017experiments}). 
All experiments employ TiO$_2$ as the photocatalyst, at standard atmospheric pressure, with fixed physical constants (diffusivity, viscosity). 
The systems differ in chemical properties, reactor design and fitted kinetic parameters (Table~\ref{ref:parameters_1}). 
CH$_4$ exhibits the highest adsorption constant ($K = 4.6\times10^{-5}$ m$^3$/mol), while NO$_\text{x}$ ($3.9 \times 10^{-5}$ m$^3$/mol) and HCHO ($1.1 \times 10^{-5}$ m$^3$/mol) have lower values, reflecting weaker adsorption onto the catalyst surface. 
NO$_\text{x}$ has the highest rate constant ($k' = 5.22$ mol$\cdot$m$^{2(a-1)}$/(s$\cdot$W$^{a}$)), followed by HCHO ($1.48$ mol$\cdot$m$^{2(a-1)}$/(s$\cdot$W$^{a}$)) and CH$_4$ (0.084 mol$\cdot$m$^{2(a-1)}$/(s$\cdot$W$^{a}$)), indicating that NO$_\text{x}$ reacts most rapidly once adsorbed on the catalyst surface, although a direct comparison is complicated by the units of $k'$ depending on $a$, which differs across pollutants. 
Light-intensity exponents vary across experiments (0.23 for NO$_\text{x}$, 0.38 for HCHO, 0.21 for CH$_4$). 
Values of $a<0.5$ indicate that the systems operate in a regime where mass-transport limitation contributes to the photon-utilisation efficiency, in addition to electron–hole recombination losses that are typically associated with $a\approx0.5$ at moderate-to-high irradiance~\cite{malayeri2019modeling}.

For all PCO systems, conversion efficiency decreases non-linearly with concentration due to Langmuir--Hinshelwood kinetics \eqref{eq:LH}, with the steepest decline observed for NO$_\text{x}$ (Figure~\ref{fig:removal_map}d).  
CH$_4$ and HCHO show more gradual declines in conversion efficiency due to the combined effects of adsorption and reaction at the catalyst surface (Figure~\ref{fig:removal_map}a,\,g). 
The conversion efficiency increases with light intensity and begins to saturate at high irradiance for CH$_4$ and NO$_\text{x}$ (Figure~\ref{fig:removal_map}b,\,e).
In contrast, HCHO operates at lower light intensities and exhibits less saturation within the investigated range (Figure~\ref{fig:removal_map}h). 
\update{Agreement between model predictions and experimental data is good across all cases, with root-mean-square errors of 1.02\% for CH$_4$, 2.46\% for NO$_\text{x}$ and 0.36\% for HCHO.
Deviations are attributable to measurement uncertainty, humidity effects not captured in the dry CH$_4$ experiments, and simplifications in the assumed reactor geometry.}
The asymptotic solutions accurately reproduce the numerical results in their respective regions of validity, providing a closed-form alternative to numerical simulation. 
Deviations from model predictions may be attributable to low-concentration measurement errors, humidity or flow effects, depending on the specific experimental setup. 
All experimental realisations are illustrated in non-dimensional $\textrm{Da}$--$\textrm{Pe}$ space (white symbols in Figure~\ref{fig:removal_map}c, f, i), where the magnitudes of $\textrm{Da}$ and $\textrm{Pe}$ place the CH$_4$, NO$_\text{x}$ and HCHO experiments on the boundary between low and high removal. 
Asymptotic solutions in this region (Appendix~\ref{sec:Region I and II: small Pe}) predict where $\eta = 10$\% accurately (dashed lines in Figure~\ref{fig:removal_map}c, f, i), providing guidance for reactor design to achieve high conversion, i.e., $\textrm{Da} \gg \textrm{Pe}$. 

\subsection{Apparent quantum yield} \label{subsec:Apparent quantum yield}

\begin{figure}
    \centering
    (a) \hfill (b) \hspace{2cm} \hfill \hfill \hfill \\    \includegraphics[width=\textwidth,trim={0.2cm 0 0.1cm 0},clip]{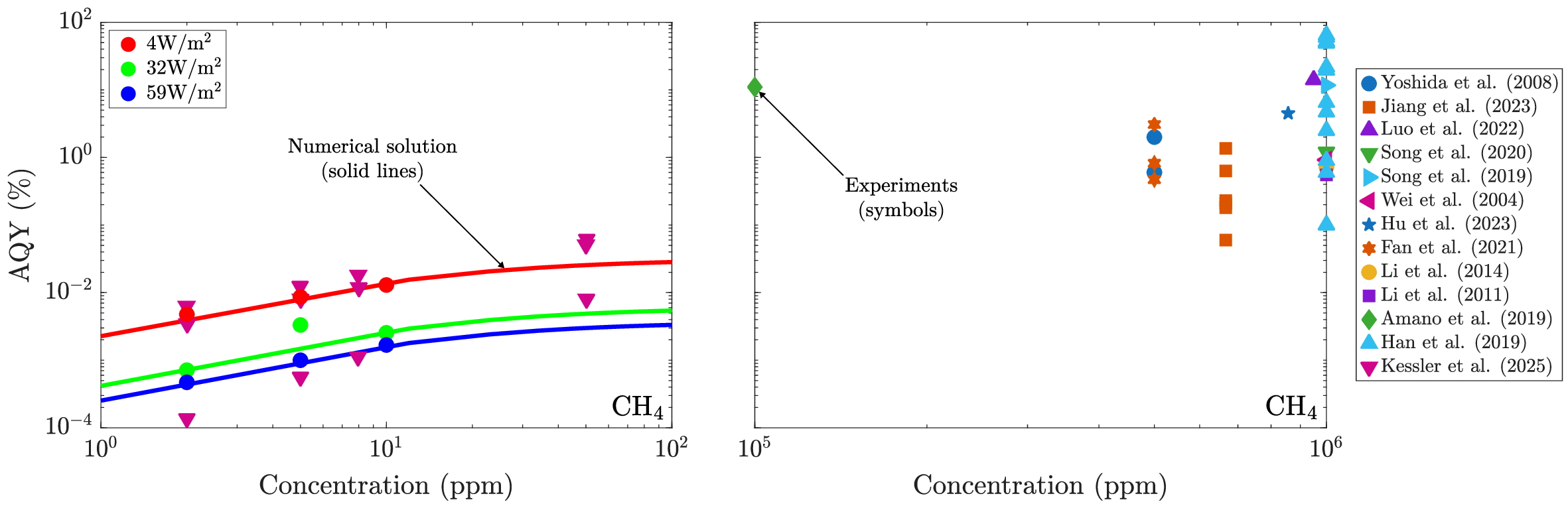}
    \caption{Apparent quantum yield (\%) for CH$_4$ oxidation for varying inlet concentrations (ppm) and light intensities (W/m$^2$). 
    Symbols denote experimental measurements and solid lines show model predictions using the quantities in Table~\ref{ref:parameters_1}. 
    \update{In (a) we show low-concentration data ($c_{\text{in}}\in[10^0\,\text{ppm},\,10^2\,\text{ppm}]$) and in (b) we show high-concentration data ($c_{\text{in}}\in[10^5\,\text{ppm},\,10^6\,\text{ppm}]$)~\cite{kessler2026humidity, Han2019, Luo2021, Fan2021, Yoshida2008, Jiang2023, Song2020, Hu2023, Li2015, Song2018, Wei2004a, Wei2004b, Li2011}.} 
    }
    \label{fig:AQY}
\end{figure}

To assess PCO performance compared to the literature, we compute the AQY (\%) of CH$_4$ oxidation across the range of concentrations (2 to 10 ppm) and UV-C light intensities (4 to 59 W/m$^2$) considered in Section~\ref{sec:Pollutant removal under variable conditions}. 
The AQY represents the ratio of reacted CH$_4$ molecules to incident photons and quantifies photon utilisation efficiency independently of the experimental setup, whereas $\eta$ reflects performance for a given experimental setup.
For each experiment, AQY was calculated from the measured reaction rate and incident photon flux as described in Section~\ref{sec:Experimental Conditions}. 
We evaluate the AQY for (i) the present experiments, (ii) other recent low-concentration photocatalyst experiments~\cite{kessler2026humidity} and (iii) the higher-concentration dataset compiled in Tsopelakou et al.~\cite{tsopelakou2024exploring}. 
The results are shown in Figure~\ref{fig:AQY}.

In Figure~\ref{fig:AQY}a, the AQY increases with increasing CH$_4$ concentration ($4.8\times10^{-3}$ to 0.013\% for 2 to 10 ppm and 4 W/m$^2$) and decreases with increasing light intensity ($4.8\times10^{-3}$ to $4.7\times10^{-4}$\% for 2 ppm and 4 to 59 W/m$^2$). 
\update{Our results are consistent with Kessler et al.~\cite{kessler2026humidity}, who recently reported AQYs of $10^{-4}$--$10^{-2}$\% for CH$_4$ at ppm-level concentrations, suggesting that lower AQYs are intrinsic to dilute-CH$_4$ photocatalysis rather than an artefact of catalyst or reactor choice.
In the high-concentration regime (Figure~\ref{fig:AQY}b), recent studies~\cite{Luo2021, Fan2021, Jiang2023, Song2020, Hu2023} achieve higher AQYs than the low-concentration systems.}
For the present experiments, the increase with concentration is most pronounced at lower concentrations ($< 10$ ppm), with the AQY approaching a plateau at higher concentrations ($> 10$ ppm). 
Increasing the light intensity enhances the conversion efficiency (Figure~\ref{fig:removal_map}) but reduces the AQY, reflecting a sub-linear dependence on light intensity ($a = 0.21$, Table~\ref{ref:parameters_1}).
Although higher photon flux accelerates the reaction rate, the reaction rate grows more slowly than the photon supply, resulting in a reduced AQY.
At low CH$_4$ concentrations, the surface coverage is small, so adding more CH$_4$ molecules substantially increases reaction rates, leading to a steep rise in AQY.
As the CH$_4$ concentration increases further, surface sites approach saturation, and the incremental gain in rate per additional CH$_4$ molecule diminishes, causing AQY to level off.
The model reproduces the trends of the present CH$_4$-experimental data in the left panel of Figure~\ref{fig:AQY}. 

\subsection{Concentration field in the reactor} \label{sec:Flow field and concentration distribution}

\begin{figure}[t!]
\centering
(a) \hfill (b) \hfill (c) \hfill \hfill \hfill \\
\includegraphics[width=0.33\textwidth]{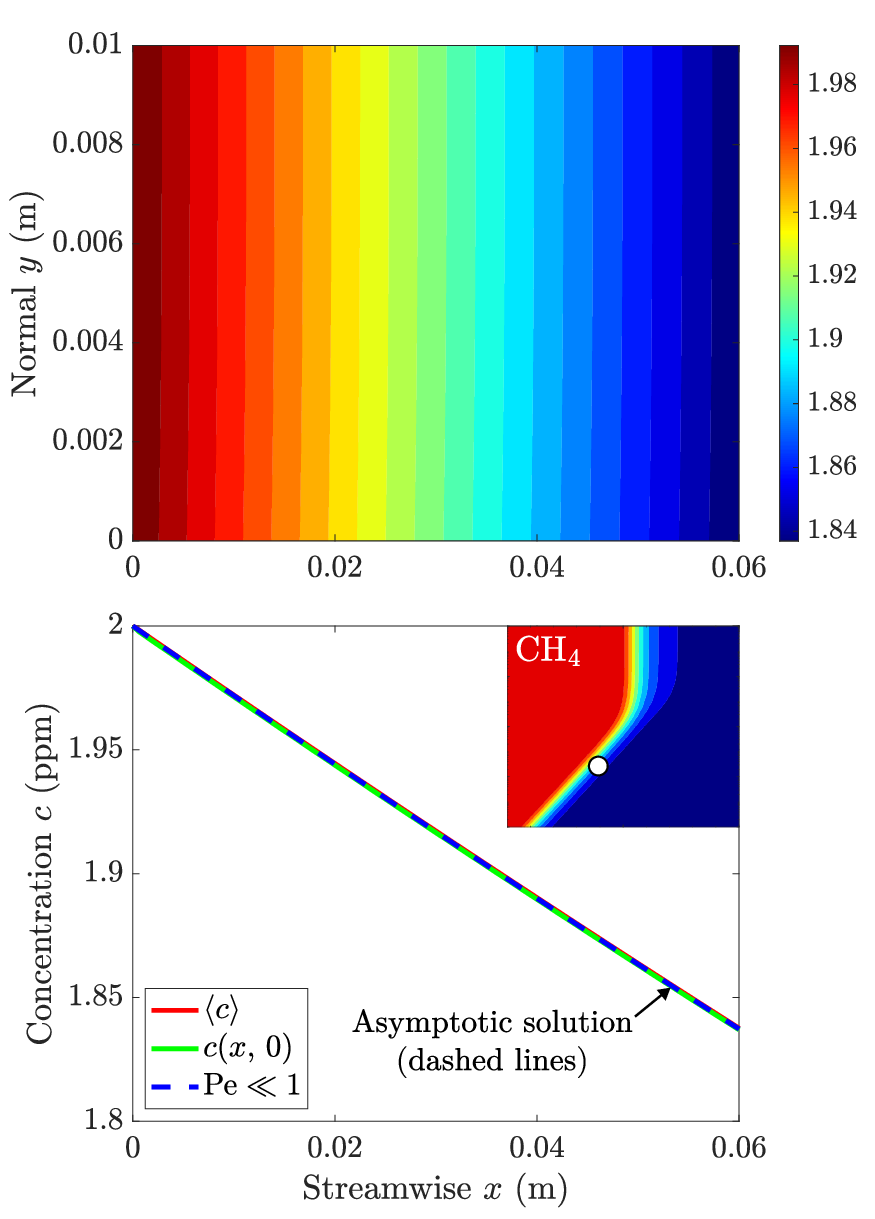}\hfill\includegraphics[width=0.33\textwidth]{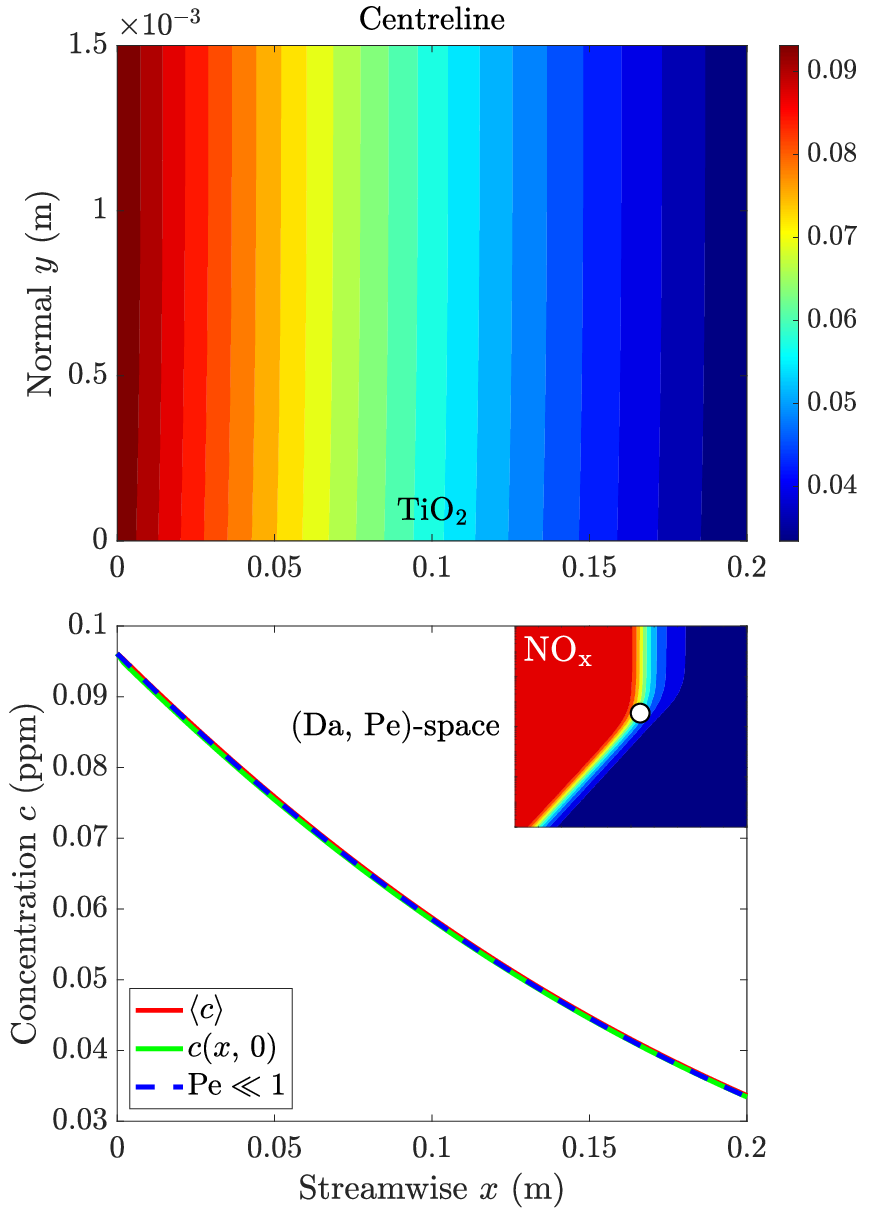}\hfill\includegraphics[width=0.33\textwidth]{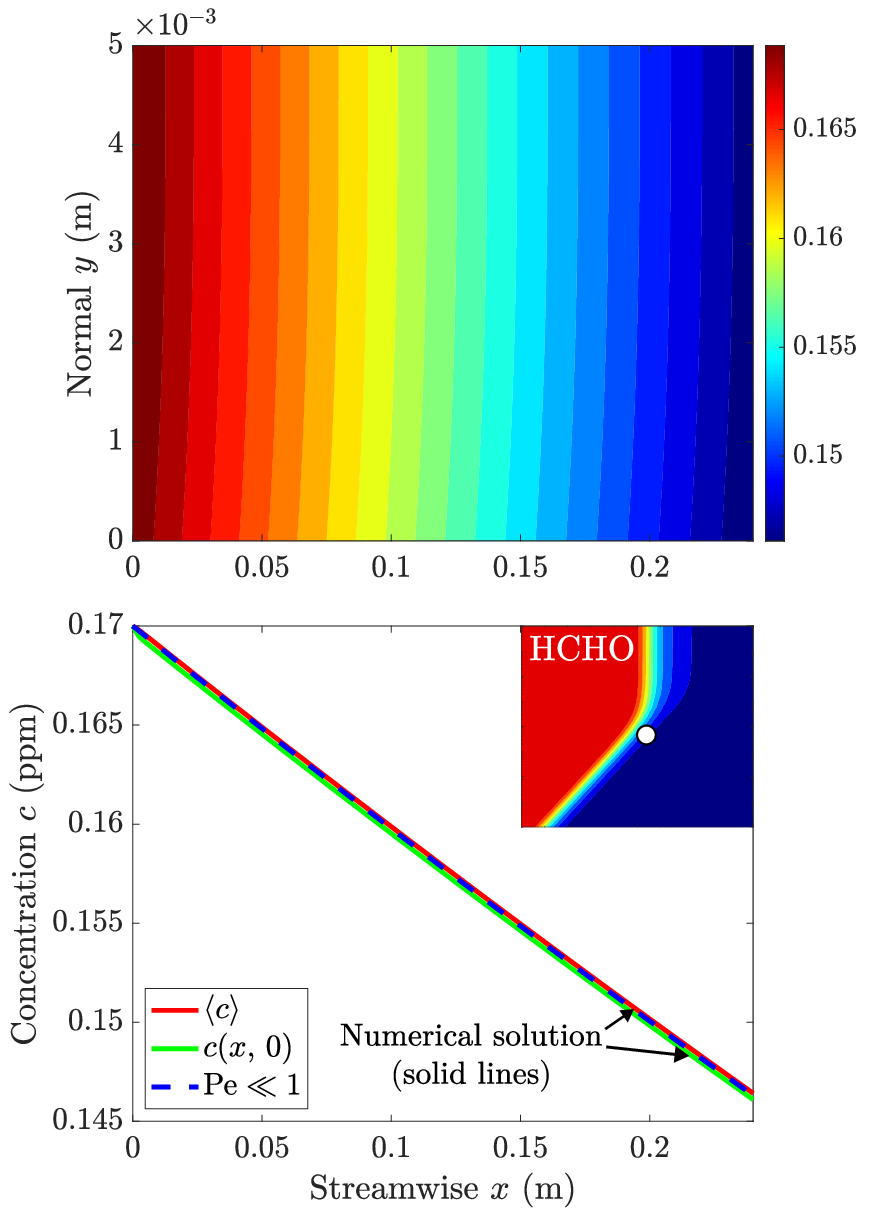}\hfill\hfill\hfill\\
\caption{Contours of the concentration field $c(x,\,y)$ for maximum removal in the laboratory-scale system. 
(a) CH$_4$ ($\textrm{Da} = 0.027$, $\textrm{Pe} = 0.085$ and $\beta = 0.17$; Section~\ref{sec:Experiments}), (b) NO$_\text{x}$ ($\textrm{Da} = 3.45$, $\textrm{Pe} = 2.50$ and $\beta = 0.54$; Ballari et al.~\cite{ballari2010modelling}) and (c) HCHO ($\textrm{Da} = 0.46$, $\textrm{Pe} = 2.50$ and $\beta = 0.26$; Yu et al.~\cite{yu2017experiments}). 
The inset shows the location in $(\mathrm{Da}, \mathrm{Pe})$ space and contours of $\eta$ from 0\% (blue) to 100\% (red).
Solid lines represent the numerical solution of the model \eqref{eq:advdiff}--\eqref{eq:topwall}, while dashed lines indicate the asymptotic solution in the small-$\textrm{Pe}$ limit \eqref{eq:rem_eff_small_Pe}. 
}
\label{fig:concentration_profile_reactor}
\end{figure}

To examine pollutant transport in the different PCO reactors, we compute steady-state 2D concentration fields for CH$_4$, NO$_\text{x}$ and HCHO. 
Pollutant transport occurs predominantly via advection in the streamwise direction and diffusion in the wall-normal direction, with photocatalytic reaction at the bottom surface ($y = 0$) and a no-flux condition at the top surface ($y = 2H$). 
Figure~\ref{fig:concentration_profile_reactor} shows the concentration field for CH$_4$ (panel a), NO$_\text{x}$ (b) and HCHO (c) under maximum removal conditions, as identified in Figure~\ref{fig:removal_map}. 
The streamwise concentration gradients result from the combined effects of surface kinetics, flow rate, light intensity and pollutant adsorption. 

The CH$_4$ and HCHO systems exhibit approximately linear concentration profiles along $x$ (Figure~\ref{fig:concentration_profile_reactor}a,\,c) with weak axial gradients. 
This is consistent with $\mathrm{Da}/\mathrm{Pe} \ll 1$, for which the concentration varies slowly along the reactor and the influence of Langmuir--Hinshelwood nonlinearity (through $\beta$) is negligible. 
In contrast, the NO$_\text{x}$ system shows a sharper inlet decrease followed by slight downstream flattening (Figure~\ref{fig:concentration_profile_reactor}b), consistent with $\mathrm{Da}/\mathrm{Pe} = O(1)$.
In this regime, adsorption effects (through $\beta$) begin to influence the kinetics, leading to the observed curvature. 
In all cases, $\mathrm{Pe} \ll 1$, and cross-channel concentration gradients remain weak, indicating that transverse diffusion is sufficiently rapid to maintain near-uniformity in $y$ and strong coupling between reaction and cross-channel mass transport. 

\subsection{Concentration field in ventilation applications} \label{sec:Full-duct simulation and scale-up}

\begin{figure}[t!]
\centering
(a) \hfill (b) \hfill (c) \hfill \hfill \hfill \\
\includegraphics[width=0.33\textwidth]{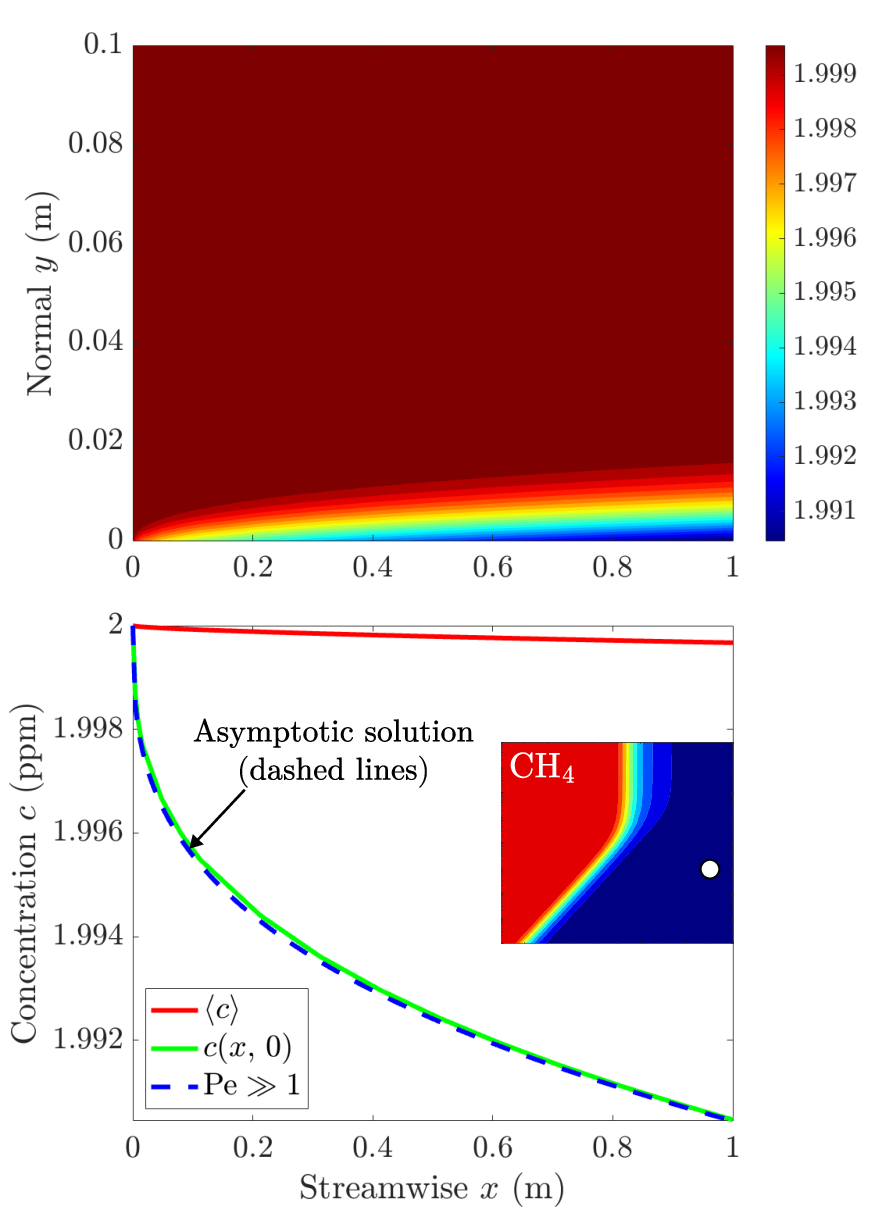}\hfill\includegraphics[width=0.33\textwidth]{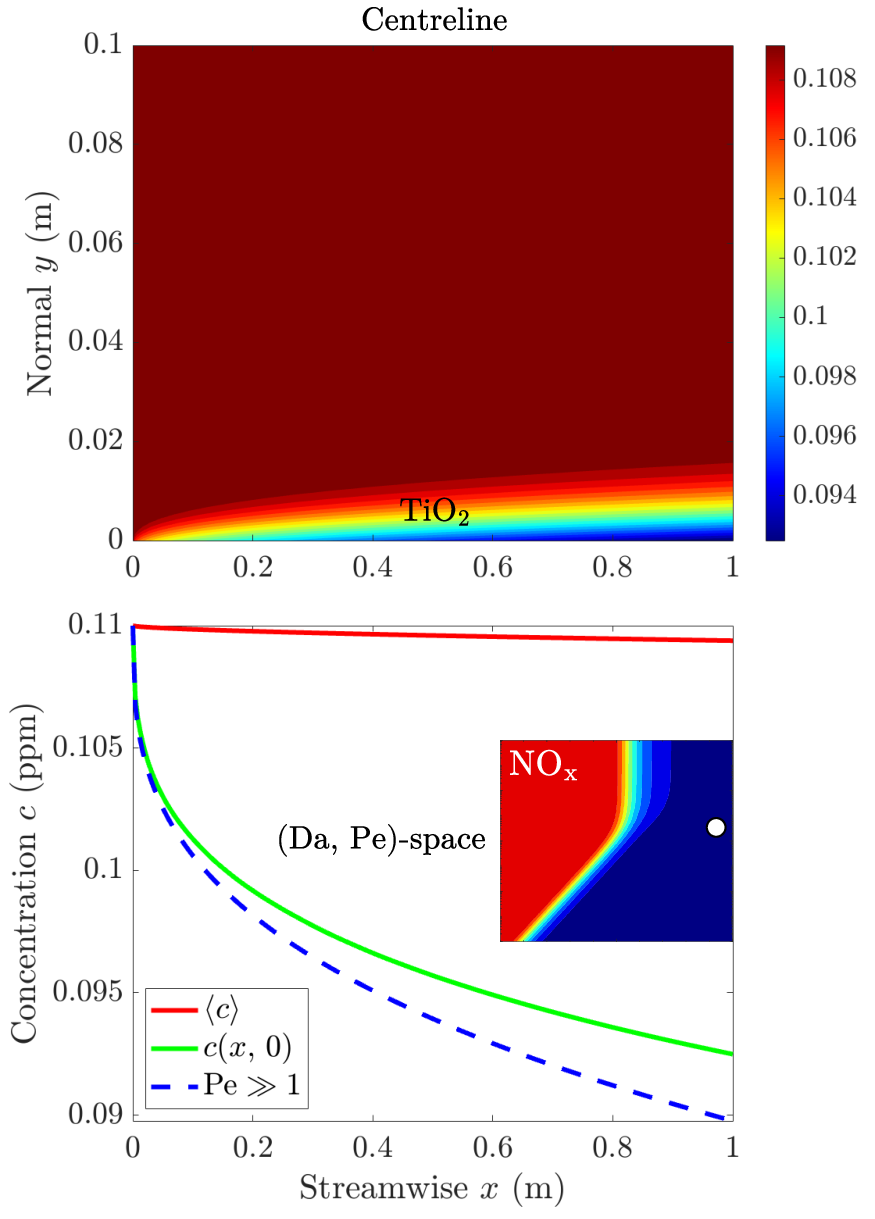}\hfill\includegraphics[width=0.33\textwidth]{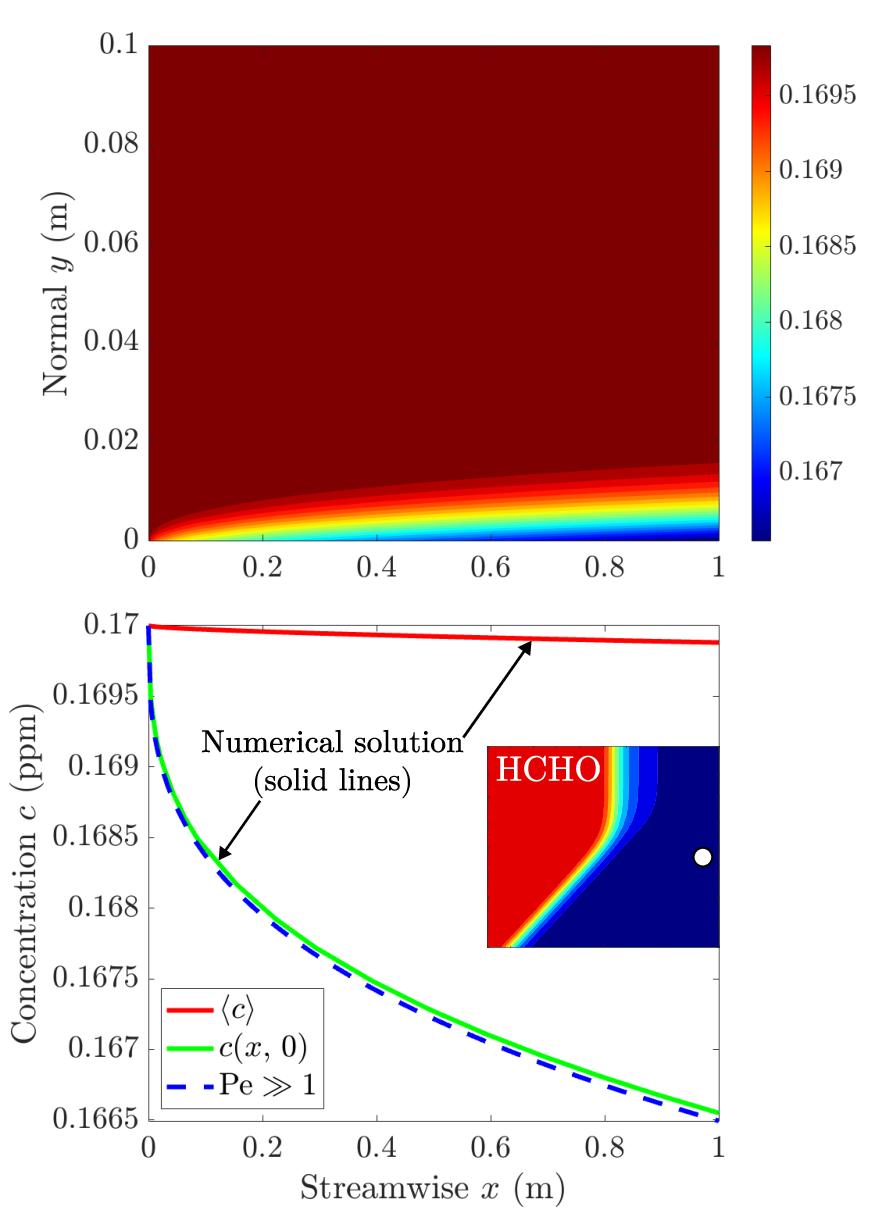}\hfill\hfill\hfill\\
\caption{Contours of the concentration field $c(x,\,y)$ for maximum removal in the duct-scale system. 
(a) CH$_4$ ($\textrm{Da} = 0.090$, $\textrm{Pe} = 1\times10^4$ and $\beta = 0.17$; Section~\ref{sec:Experiments}), (b) NO$_\text{x}$ ($\textrm{Da} = 3.46$, $\textrm{Pe} = 2\times10^4$ and $\beta = 0.57$; Ballari et al.~\cite{ballari2010modelling}) and (c) HCHO ($\textrm{Da} = 0.39$, $\textrm{Pe} = 2\times10^4$ and $\beta = 0.26$; Yu et al.~\cite{yu2017experiments}). 
The inset shows the location in $(\mathrm{Da}, \mathrm{Pe})$ space and contours of $\eta$ from 0\% (blue) to 100\% (red).
Solid lines represent the numerical solution of the model \eqref{eq:advdiff}--\eqref{eq:topwall}, while dashed lines indicate the asymptotic solution in the large-$\textrm{Pe}$ limit \eqref{eq:removal_eff_large_Da_final}. 
}
\label{fig:concentration_profile_duct}
\end{figure}

To evaluate the scalability of PCO for indoor air treatment, we extend the laboratory-scale model from Section~\ref{sec:Model} to a duct geometry representative of ventilation systems. 
The duct is assumed to have a rectangular cross-section with a height of 0.2 m, a width of 0.2 m and a 1 m long test section, uniformly illuminated with UV-C light. 
Air flows at a rate of 0.08 m$^3$/s, corresponding to an average velocity of approximately 2 m/s. 
These operating conditions enhance advective transport and reduce residence time relative to the laboratory-scale setup, resulting in elevated Reynolds numbers and the formation of a concentration boundary layer near the catalyst surface. 
We assume that the bottom wall of the duct is coated with TiO$_2$, while other walls are treated as no-flux boundaries. 

Figure~\ref{fig:concentration_profile_duct} presents the simulated concentration fields in the duct under the high-removal conditions identified in Figure~\ref{fig:removal_map}. 
In contrast to the laboratory reactor (Figure~\ref{fig:concentration_profile_reactor}), the duct-scale concentration field exhibits steep wall-normal gradients near the catalyst, while concentrations remain nearly uniform outside the boundary layer, indicating that the system is mass-transfer-limited. 
Under these conditions, removal efficiencies are low: 0.010\% to 0.017\% for CH$_4$, 0.56\% for NO$_\text{x}$ and 0.039\% to 0.070\% for HCHO. 
Table~\ref{tab:scaling_up} summarises these results, showing the estimated mass removed for CH$_4$, NO$_\text{x}$ and HCHO, together with the associated experimental parameters from Section~\ref{sec:Experiments}, Ballari et al.~\cite{ballari2010modelling} and Yu et al.~\cite{yu2017experiments}. 
The annual removal for the assumed test section is modest, ranging from $2.1\times10^{-7}$ to $2.9\times10^{-5}$ tonnes per pollutant per year across CH$_4$, NO$_\text{x}$ and HCHO, illustrating the large surface area required to achieve large removal rates in ventilation systems~\cite{tomlinson2025harnessing}. 

\subsection{Design implications}

\update{In the laboratory reactor ($\mathrm{Da} = 0.016$--0.027, $\mathrm{Pe} = 0.085$), the system is in the $\mathrm{Da} \ll 1$ and $\mathrm{Pe} \ll 1$ regime, where Appendix A.1 shows that $\eta \approx \mathrm{Da}/\mathrm{Pe}$.  
Hence, the conversion efficiency increases with $I^a$, $k'$ and $K$ and decreases with $\langle u \rangle$.  
At duct scale ($\mathrm{Da} \approx 0.09$, $\mathrm{Pe} \approx 10^4$), Appendix A.2 gives $\eta \approx 0.43\,\mathrm{Da}/\mathrm{Pe}^{2/3}$ and we are in the reaction-limited limit ($\mathrm{Da} \ll 1$). 
Hence, in this regime, an increase in light intensity and catalyst activity, or decrease in flow velocity (increasing residence time), yields an improvement in conversion. 
However, for $\mathrm{Da} \gg 1$ we have a transport-limited ceiling of $\eta \approx 0.58/10^{4/3} \approx 2.7\%$, which is independent of light intensity and catalyst activity.  
Reaching this ceiling requires $\mathrm{Da} \gg 1$, i.e., an increase in $I^a$, $k'$ or $K$ of approximately two orders of magnitude from the TiO$_2$ baseline. 
This analysis shows that catalyst improvement must be pursued up to a point; catalyst gains alone cannot exceed the 2.6\% ceiling at the current ventilation geometry and flow rate. 
Therefore, reducing the flow velocity is the main way to raise this ceiling and achieve considerably higher conversions.}

\subsection{Climate benefit for UV light in ventilation applications} \label{sec:Climate benefit for UV light}

The PCO system described in Section~\ref{sec:Full-duct simulation and scale-up} relies on UV-C irradiation to activate TiO$_2$ surfaces and oxidise CH$_4$. 
This approach achieves high CH$_4$ conversion (Section~\ref{sec:Pollutant removal under variable conditions}); however, it requires electrical energy for UV illumination, resulting in associated CO$_2$e emissions. 
In addition, the embodied carbon of the TiO$_2$ catalyst contributes to the climate impact and must be included in the assessment~\cite{RioTinto2020Climate}. 
The use of metal-free TiO$_2$ results in lower embodied carbon compared to metal-loaded catalysts~\cite{marina2026evaluating}. 
The net climate impact of CH$_4$ PCO is assessed for a duct-scale system using the methodology detailed in Appendix~\ref{sec:Methodology for photocatalyst}. 
Note that the catalytic performance reported here represents short-term activity under controlled laboratory conditions. 
The net CO$_2$e emissions analysis assumes stable catalyst performance and does not account for potential deactivation or replacement requirements. 
Evaluation of catalyst durability, structural evolution and lifetime performance in realistic environments is therefore an important direction for future work.

\begin{table}[t!]
\centering
\caption{Summary of the predicted removal rates and climate benefit metrics for scaling up photocatalytic studies on CH$_4$, HCHO and NO$_\text{x}$ removal using TiO$_2$ to a ventilation system of height 0.2 m, width 0.2 m and 1 m length. 
The flow rate is 0.08 m$^3$/s and the UV-C lamp is 25 W, which provides the range of light intensities in the laboratory experiments.
}
\vspace{.25cm}
\renewcommand{\arraystretch}{1.2}
\small
\begin{tabular}{|l|c|c|c|}
\hline
Parameter & Ballari et al. (2010) & Yu et al. (2017) & Present \\
\hline
\multicolumn{4}{|c|}{Removal quantities} \\
\hline
Concentration (ppm) & 0.11 to 1.09 & 0.17 to 0.9 & 2 to 10 \\
Removal efficiency (\%) & 0.5565 & 0.0389 to 0.0702 & 0.0095 to 0.0165 \\
Max. removal rate ($\times 10^{-3}$ ppm/s) & 0.612 to 6.07 & 0.066 to 0.632 & 0.190 to 8.25 \\
Max. mass removed ($\times 10^{-5}$ t/y) & 0.290 to 2.88 & 0.021 to 0.196 & 0.031 to 1.36 \\ 
\hline
\multicolumn{4}{|c|}{Climate benefit for UV light} \\
\hline
CO$_2$e removal rate ($\times 10^{-3}$ tCO$_2$e/y) 
& -- & -- & 0.026 to 1.14 \\
\hline
Net CO$_2$e w/ UV ($\times 10^{-2}$ tCO$_2$e/y) 
& -- & -- & 4.27 to 4.38 \\
\hline
Net CO$_2$e w/o UV ($\times 10^{-3}$ tCO$_2$e/y) 
& -- & -- & $-1.13$ to $-0.014$ \\
\hline
\end{tabular}
\label{tab:scaling_up}
\end{table}

Each system is assumed to operate with a 25 W UV lamp, which we assume provides the maximum irradiance over the catalyst surface, running continuously 24 h/day throughout the year (see Table~\ref{tab:parameters_pco}). 
Under these operating conditions, the system consumes around 219 kWh/y of electrical energy, corresponding to 0.044 tCO$_2$/y of emissions for a grid electricity intensity of 200 gCO$_2$/kWh. 
The embodied carbon of the catalyst is based on a TiO$_2$ mass of 3.2 g and an emission factor of 4 gCO$_2$e/gTiO$_2$~\cite{RioTinto2020Climate}. 
Assuming an annual replacement cycle for the catalyst (an upper-bound assumption pending lifetime studies), this yields an annual contribution of $1.3\times10^{-5}$ tCO$_2$e/y. 
The pollutant removal efficiency from Section~\ref{sec:Full-duct simulation and scale-up} was applied to the inlet concentration to calculate the pollutant removal per unit volume, which was then converted to a removal rate using the ventilation flow rate, as summarised in Table~\ref{tab:scaling_up}.
Under the specified operating conditions, the system is predicted to remove approximately $3.1\times10^{-7}$ to $1.4\times10^{-5}$ tCH$_4$/y. 
Considering the 20-year global warming potential of CH$_4$ (GWP$_{20}$ = 84~\cite{kumar2021climate}), the annual CH$_4$ removal corresponds to a climate benefit of approximately $2.6\times10^{-5}$ to $1.1\times10^{-3}$ tCO$_2$e/y. 
When the CO$_2$e emissions from UV electricity consumption and catalyst embodied carbon are included, the net climate effect is predicted to be a net positive emission of approximately $4.3\times10^{-2}$ to $4.4\times10^{-2}$ tCO$_2$e/y, indicating that the energy cost of UV operation outweighs the climate benefit of CH$_4$ removal in this configuration. 
However, if the UV energy is assumed to be already available (e.g., in a pre-existing disinfection systems), the net climate effect is predicted to be approximately $-1.1\times10^{-3}$ to $-1.4\times10^{-5}$ tCO$_2$e/y, indicating that the system would be effectively removing CO$_2$e.
Overall, this calculation demonstrates that duct-scale TiO$_2$-based PCO may provide a measurable net climate benefit when leveraging pre-existing UV irradiation, even when accounting for catalyst embodied carbon. 

\update{For a commercial building HVAC system, assuming the entire ductwork (e.g., 100 m of catalyst-coated duct) operates under pre-existing UV-C irradiation, the predicted maximum annual CH$_4$ removal scales to approximately $1.4\times10^{-3}$ tCH$_4$/y ($0.12$~tCO$_2$e/y). 
Ventilation and lighting each account for over 14\% of commercial building electricity use~\cite{eia2022cbecshighlights}  
At an average electricity use of 203 MWh/y per building~\cite{eia2022cbecshighlights}, this gives Scope 2 emissions of 5.7 tCO$_2$e/y at 200 gCO$_2$/kWh.  
The maximum PCO removal benefit for 100 m of catalyst-coated duct is therefore 2.1\% of this.} 
\update{However, UV-C illuminated ductwork is not yet widespread, though installation of UV-C disinfection systems in commercial HVAC has been growing since 2020~\cite{thornton2022impact}; where currently installed, light intensities are optimised for microbial disinfection rather than photocatalysis, so the climate-benefit estimates above should be interpreted as an upper bound contingent on this infrastructure becoming more common.}
\update{Because atmospheric CH$_4$ is highly dilute, this removal is necessarily small relative to total building-scale emissions regardless of catalyst performance; it is also useful to weigh this benefit against the CO$_2$e cost of operating the UV-C source itself, as shown above, and meaningful climate impact would require deployment far beyond a single building, consistent with the large-area estimates in Tomlinson et al.~\cite{tomlinson2025harnessing}.}

\section{Discussion} \label{sec:Discussion}

The modelling framework developed here provides a description of PCO for CH$_4$, NO$_\text{x}$ and HCHO. 
Comparison of the model with our CH$_4$ experiments and literature data for NO$_\text{x}$ and HCHO (Figure~\ref{fig:removal_map}) shows that the fitted kinetic parameters, the rate coefficient $k'$, adsorption constant $K$ and light exponent $a$, reproduce measured removal efficiencies and differentiate pollutants with fast surface kinetics from those limited by weak adsorption. 
For CH$_4$, strong adsorption increases surface coverage and yields high conversion despite slower surface kinetics. 
In contrast, NO$_\text{x}$ oxidation is governed by faster surface kinetics even though adsorption is weaker. 
HCHO exhibits the lowest activity because its adsorption constant and rate coefficient are small, resulting in limited surface coverage and slow reaction. 
These differences explain the pollutant-specific removal trends predicted by the model and show how surface coverage and saturation constrain PCO performance at environmentally relevant concentrations. 
Extending the lab-scale results~\cite{yu2017experiments, ballari2010modelling} to duct geometries quantifies how reaction and transport mechanisms are affected under ventilation flow conditions. 
Thin diffusive boundary layers near the walls confine reactions to this region, limiting catalyst utilisation and resulting in removal efficiencies $<0.1\%$ for all pollutants, even under high irradiance (Figure~\ref{fig:concentration_profile_duct}). 
For CH$_4$, our analysis predicts that duct-scale operation can achieve a net climate benefit (net-negative CO$_2$e emissions) when UV energy is already available (e.g., in pre-existing disinfection systems), even after accounting for the embodied carbon of the catalyst.
This benefit is primarily enabled by the use of metal-free TiO$_2$ coatings, which minimise material-related emissions compared to more complex catalyst formulations~\cite{marina2026evaluating}.

\update{The present model assumes laminar flow as an approximation that yields an analytical estimate of mass transfer. 
In practice, ventilation-scale operating conditions may fall in the turbulent regime, which would enhance mass transfer to the catalyst surface relative to the laminar case~\cite{frank2017incropera}. 
Surface roughness, mixing baffles or other modifications could also be used to enhance mass transfer. 
However, they would increase pressure drop and fan energy consumption, offsetting part of the climate benefit quantified in Section~\ref{sec:Climate benefit for UV light}. 
A promising configuration is photocatalytic coating of HVAC filters~\cite{Tomlinson2026}: filters are engineered to maximise fluid--surface interaction, and the resulting higher surface area and longer residence time could increase the mass transfer by several orders of magnitude compared to a duct-wall coating.} 
\update{
Furthermore, the concentration boundary-layer analysis presented here for ventilation flow has analogues for other applications, such as TiO$_2$-coated building facades, roads and rooftops. 
In the large-$\textrm{Pe}$ limit (Appendix~\ref{ref:asym_4}), the near-wall boundary-layer solution depends only on the local streamwise distance and $\textrm{Pe}$, not on the rest of the geometry, so the same analytical estimates should apply to an external flow over a building facade or rooftop. 
For a characteristic wind speed of 4 m/s and a building length scale of 30 m, this gives $\mathrm{Pe} \approx 6.7\times10^6$; using the NO$_\text{x}$ kinetic parameters fitted to Ballari et al. (Table~\ref{ref:parameters_1}) gives $\mathrm{Da} \approx 576$, placing the system firmly in the transport-limited regime. 
The resulting mass-transfer-limited removal efficiency, $\eta \approx 0.58/\mathrm{Pe}^{1/3} \approx 0.31\%$, is of a similar order of magnitude to the 2\% NO$_\text{x}$ reduction reported by Russell et al.~\cite{Russell2021Review} for outdoor TiO$_2$ coatings. 
Our model does not, however, include surface roughness or turbulence, and resolving turbulent transport requires numerical simulation beyond the scope of this manuscript.  
Extending this work to include surface roughness and turbulent flow regimes, and quantifying the associated pressure-drop penalty in ventilation applications, are left for future work.} 

Several other limitations require further investigation. 
The reported metrics isolate the catalytic contribution and do not include the bulk oxidation of CH$_4$ by reactive oxygen species and OH$\cdot$ generated by the UV-C lamp~\cite{krogsboll2025efficient, Pennacchio2024, iversen2025mitigation}; total conversion efficiencies and the bulk reaction mechanism will be quantified in future work.
The analysis assumes uniform irradiance, neglecting photon scattering and catalyst shadowing, which can create spatial variations in surface reactivity~\cite{malayeri2019modeling}. 
Catalyst ageing, deactivation and fouling are not included in the model and may reduce long-term performance. 
\update{TiO$_2$-based photocatalysts can deactivate through accumulation of partially oxidised intermediates~\cite{pitchai1986}, poisoning by trace contaminants~\cite{zhou2025}, or sintering under prolonged UV exposure~\cite{argyle2015heterogeneous}.
Catalyst regeneration has been reported to restore activity in some cases~\cite{Anekwe2025}, but long-term durability under realistic ventilation conditions remains an open question.
Mitigation strategies for these effects include catalyst immobilisation and upstream filtration.}
Thus, the reported metrics reflect intrinsic catalytic behaviour under idealised conditions rather than long-term field performance.
Avenues for future work include catalyst durability under humid and contaminant-containing feeds that are more representative of environmental methane sources.
The potential formation of secondary species and partially oxidised by-products~\cite{pitchai1986}, has not been quantified and requires targeted experimental investigation. 
Addressing these factors is essential for assessing real-world PCO performance. 
\update{The PCO of CH$_4$ is expected to proceed through successive C--H bond activation steps, producing intermediate species including methanol (CH$_3$OH), formaldehyde (HCHO), formic acid (HCOOH) and ultimately CO$_2$ and H$_2$O~\cite{pitchai1986}.
In the present experiments, the absence of additional FID peaks in the chromatograms supports the assumption of complete oxidation to CO$_2$; however, trace intermediates below the detection limit cannot be ruled out and their quantification is an important direction for future work.}

Despite these limitations, the framework provides a foundation for the optimisation of PCO ventilation technologies. 
Integrating experimentally validated kinetics with detailed transport modelling allows quantitative evaluation of design modifications, including increased catalyst area, enhanced UV irradiance, or integration with complementary air-cleaning methods. 
Alternative ventilation system designs could also be explored~\cite{Tomlinson2026}. 
For example, coating ventilation filters may enhance mass transfer and pollutant removal efficiency, as pollutants would be transported through the catalytic medium rather than merely over a coated surface. 
Future work could integrate sensor-driven control strategies to adjust UV irradiation or airflow dynamically based on real-time pollutant concentrations. 
A comprehensive life-cycle assessment will also be necessary to quantify the long-term environmental and operational sustainability of PCO systems. 

In summary, our experimental and modelling results investigate TiO$_2$-based PCO as a potential strategy for improving indoor air quality and, in the case of CH$_4$, predict that it could potentially achieve a net climate benefit if integrated into existing UV-illuminated infrastructure. 
\update{However, the net climate benefit predicted under the assumption of pre-existing UV-C infrastructure is negligible relative to building-scale CO$_2$e emissions. 
Meaningful atmospheric impact would require large-scale deployment across millions of square metres of catalytic surface, as quantified in~\cite{tomlinson2025harnessing}, which faces substantial economic barriers~\cite{randall2024cost, hickey2024economics}.
We further note that PCO could, in principle, target multiple pollutants simultaneously using the same UV-C infrastructure, which would improve both the health and climate case for deployment; quantifying this multi-pollutant co-benefit is left for future work.} 

\vspace{0.5cm}

\noindent\textbf{Acknowledgments:} We acknowledge Grantham Foundation for supporting this research. 
SDT acknowledges funding from Research England's E3 fund via the M$^3$4Impact program.

\appendix

\section{Asymptotic analysis} \label{sec:Asymptotic analysis}

In Section~\ref{sec:Results}, experimental results are compared with numerical and asymptotic solutions to assess model performance. 
Asymptotic solutions provide physical insight into pollutant removal and enable simple predictions in parameter regimes where thin boundary layers make full numerical solutions computationally expensive. 
In this appendix, we derive asymptotic approximations for the pollutant concentration \(\hat{c}\) and conversion efficiency \(\eta\) in three regimes. 
These regimes are defined by the relative strength of reaction \((\mathrm{Da})\), diffusion \((\mathrm{Pe})\) and adsorption \((\beta)\) compared to advection. 

\subsection{Region I and II: small Pe} \label{sec:Region I and II: small Pe}

We begin by integrating the advection--diffusion equation \eqref{eq:non_dim_1} across the channel and applying the wall flux conditions \eqref{eq:non_dim_2}, which gives
\begin{equation}
\frac{3}{2}\int_{\hat{y}=0}^1 \left(1 - (2\hat{y} - 1)^2 \right) \frac{\partial \hat{c}}{\partial \hat{x}} \, \text{d} \hat{y}  = \int_{\hat{y}=0}^1 \frac{1}{\mathrm{Pe}} \frac{\partial^2 \hat{c}}{\partial \hat{y}^2} \, \text{d} \hat{y} =  - \frac{\mathrm{Da}}{\mathrm{Pe}} \hat{c}(\hat{x}, \, 0). 
\label{eq:advection-diffusion_int}
\end{equation}
In the strong-diffusion limit, $\mathrm{Pe} \ll 1$, cross-channel concentration gradients are expected to be small, so the leading-order concentration is nearly uniform in $\hat{y}$ and depends only on $\hat{x}$, i.e., $\hat{c} \approx \hat{c}(x)$. 
Substituting this approximation into the cross-channel integrated equation \eqref{eq:advection-diffusion_int} yields the leading-order ordinary differential equation (ODE)
\begin{equation}
    \frac{\partial \hat{c}}{\partial \hat{x}} = - \frac{\mathrm{Da}}{\mathrm{Pe}} \frac{\hat{c}}{1 + \beta \hat{c}}, \quad \hat{c}(0) = 1. 
    \label{eq:leading_order}
\end{equation}
For $\beta \ll 1$, \eqref{eq:leading_order} reduces to a linear ODE, whose solution is
\begin{equation}
    \hat{c} \approx \exp\left(-\frac{\mathrm{Da} \hat{x}}{\mathrm{Pe}} \right). 
    \label{eq:leading_order_sol}
\end{equation}
Substituting \eqref{eq:leading_order_sol} into \eqref{eq::non_dim_3} gives the leading-order conversion efficiency
\begin{equation} \label{eq:rem_eff_small_Pe}
\eta \approx 1 - \exp\left(-\frac{\mathrm{Da}}{\mathrm{Pe}} \right). 
\end{equation}
For $\beta \gg 1$, integrating the ODE \eqref{eq:leading_order} with respect to $\hat{x}$ gives the implicit relation
\begin{equation}
    \ln(\hat{c}) + \beta \hat{c} = \beta -\frac{\mathrm{Da} \hat{x}}{\mathrm{Pe}}. 
    \label{eq:leading_order_sol_large_beta_0}
\end{equation}
Expanding $\hat{c} = 1 + \hat{c}_1/\beta +\dots$ in \eqref{eq:leading_order_sol_large_beta_0} and matching terms at $O(1)$ yields the first-order correction
\begin{equation}
    \hat{c}_1 = -\frac{\mathrm{Da} \hat{x}}{\mathrm{Pe}}.
\end{equation}
Therefore, the concentration field can be approximated as
\begin{equation}
    \hat{c} = 1 - \frac{\mathrm{Da} \hat{x}}{\beta \mathrm{Pe}} +\dots.
    \label{eq:leading_order_sol_large_beta}
\end{equation}
Substituting the concentration field \eqref{eq:leading_order_sol_large_beta} into \eqref{eq::non_dim_3} yields the leading-order conversion efficiency 
\begin{equation} \label{eq:rem_eff_small_Pe_large_beta}
\eta = \frac{\mathrm{Da}}{\beta \mathrm{Pe}} +\dots. 
\end{equation}

\subsection{Region III: small Da, large Pe}

When $\textrm{Pe} \gg 1$, advection dominates cross-channel diffusion and a thin concentration boundary layer develops adjacent to the reactive wall at $\hat{y} = 0$. 
Retaining the leading-order terms in the advection--diffusion equation \eqref{eq:non_dim_1} in the near-wall region $\hat{y} \ll 1$ yields
\begin{equation}
6\hat{y}\frac{\partial \hat{c}}{\partial \hat{x}}  = \frac{1}{\mathrm{Pe}} \frac{\partial^2 \hat{c}}{\partial \hat{y}^2},
\label{eq:advection-diffusion_small_y}
\end{equation}
with boundary conditions given by \eqref{eq:non_dim_2}. 
For $\mathrm{Da} \ll 1$, the concentration field is expanded as
\begin{equation}
\hat{c} = 1 + \mathrm{Da} \, \hat{c}_1(\hat{x}, \,\hat{y}) + \dots,
\label{eq:expansion_small_y}
\end{equation}
where $\hat{c}_1$ satisfies the linear boundary-layer equation \eqref{eq:advection-diffusion_small_y}. 
For $\beta \ll 1$, the wall-flux condition reduces to $\hat{c}_{1\hat{y}}(\hat{x}, \, 0)=1$, and we seek a similarity solution of the form
\begin{equation}
\hat{c}_1 = \frac{\hat{x}^{1/3}}{\textrm{Pe}^{1/3}} f(\xi), \quad \xi = \frac{\hat{y} \, \textrm{Pe}^{1/3}}{\hat{x}^{1/3}}. 
\label{eq:similarity}
\end{equation}
Substituting the asymptotic expansion \eqref{eq:expansion_small_y} and similarity form \eqref{eq:similarity} into the advection--diffusion equation \eqref{eq:advection-diffusion_small_y}, together with the boundary conditions \eqref{eq:non_dim_1}--\eqref{eq:non_dim_2}, yields the following boundary-value problem 
\begin{equation}
f'' - 2\xi f + 2 \xi^2 f'= 0, \quad f'(0) = 1, \quad f(\xi \to \infty) \to 0. 
\label{eq:similarity_ode}
\end{equation}
The solution to the boundary-value problem \eqref{eq:similarity_ode} is
\begin{equation}
f = \frac{\xi\Gamma(-1/3, \, 2\xi^3/3)}{\Gamma(-1/3)}. 
\label{eq:sim_sol}
\end{equation}
Substituting the solution \eqref{eq:sim_sol} into the similarity solution \eqref{eq:similarity} gives the first-order concentration field
\begin{equation}
\hat{c}_1 = \frac{{x}^{1/3}\xi\Gamma(-1/3, \, 2\xi^3/3)}{\textrm{Pe}^{1/3}\Gamma(-1/3)}. 
\label{eq:similarity_sol_small_Da_1}
\end{equation}
Using the definition of conversion efficiency \eqref{eq::non_dim_3}, the leading-order contribution is
\begin{equation}
\eta = -\mathrm{Da} \int_0^1 \hat{c}_1(1, \, \hat{y}) \, d\hat{y} + \dots. 
\label{eq:e_1}
\end{equation}
Substituting the similarity solution \eqref{eq:similarity_sol_small_Da_1} into \eqref{eq:e_1} and changing variables to the similarity coordinate \( \xi = \hat{y} \, \textrm{Pe}^{1/3} \) (with $\hat{x}=1$) gives
\begin{equation}
\eta = -\mathrm{Da} \int_0^1 \frac{\xi\Gamma(-1/3, \, 2\xi^3/3)}{\textrm{Pe}^{1/3}\Gamma(-1/3)} \, d\hat{y}  + \ldots = - \frac{\mathrm{Da}}{\textrm{Pe}^{1/3}} \int_0^{\infty} \frac{\xi\Gamma(-1/3, \, 2\xi^3/3)}{\textrm{Pe}^{1/3}\Gamma(-1/3)} \, d\xi  + \dots. 
\label{eq:e_2}
\end{equation}
Evaluating the integral in \eqref{eq:e_2} gives the leading-order conversion efficiency
\begin{equation}
\eta = -\frac{\mathrm{Da} (3/2)^{2/3} \Gamma(1/3)}{2 \Gamma(-1/3) \textrm{Pe}^{2/3}} + \dots \approx \frac{0.43\mathrm{Da}}{\textrm{Pe}^{2/3}}. 
\end{equation}
Note that $\Gamma(-1/3) < 0$, so the leading-order expression is positive, as required for a physical conversion efficiency.
For $\beta \gg 1$, the expansion \eqref{eq:expansion_small_y} remains valid, but the wall-flux condition \eqref{eq:non_dim_2} reduces to
\begin{align}
\left. \frac{\partial \hat{c}_1}{\partial \hat{y}} \right|_{\hat{y}=0} &= \frac{\mathrm{Da}}{\beta}. 
\label{eq:bc_2_large_beta}
\end{align}
The corresponding similarity solution, satisfying the wall-flux condition \eqref{eq:bc_2_large_beta}, is
\begin{equation}
\hat{c}_1 = \frac{\textrm{Da}\hat{x}^{1/3}}{\beta\textrm{Pe}^{1/3}} f(\xi), \quad \xi = \frac{\hat{y} \, \textrm{Pe}^{1/3}}{\hat{x}^{1/3}}. 
\label{eq:similarity_beta}
\end{equation}
Substituting the asymptotic expansion \eqref{eq:expansion_small_y} and similarity solution \eqref{eq:similarity_beta} into the advection--diffusion equation \eqref{eq:advection-diffusion_small_y}, along with the boundary conditions \eqref{eq:non_dim_2}, yields the same ODE and boundary conditions as in \eqref{eq:similarity_ode}. 
The solution of the boundary-value problem is therefore identical to \eqref{eq:sim_sol}, yielding the first-order concentration field
\begin{equation}
\hat{c}_1 = \frac{\textrm{Da}{x}^{1/3}\xi\Gamma(-1/3, \, 2\xi^3/3)}{\beta \textrm{Pe}^{1/3}\Gamma(-1/3)}, 
\label{eq:similarity_sol_small_Da_2}
\end{equation}
and the leading-order conversion efficiency
\begin{equation}
\eta = - \frac{\mathrm{Da}^2 (3/2)^{2/3} \Gamma(1/3)}{2 \beta \Gamma(-1/3) \textrm{Pe}^{2/3}} + \dots \approx \frac{0.43\mathrm{Da}}{\beta \textrm{Pe}^{2/3}}. 
\end{equation}

\subsection{Region IV: large Da, large Pe}
\label{ref:asym_4}

When $\mathrm{Da} \gg 1$, the concentration at the reactive surface $\hat{y}=0$ is nearly zero, consistent with the wall-flux condition \eqref{eq:non_dim_2}. 
We therefore consider the boundary layer near $\hat{y}=0$ once more. 
The concentration field is expanded in inverse powers of $\mathrm{Da}$ as
\begin{equation}
\hat{c} = \hat{c}_0(\hat{x},\,\hat{y}) + \frac{1}{\mathrm{Da}} \hat{c}_1(\hat{x},\,\hat{y}) + \dots,
\label{eq:expansion_large_Da}
\end{equation}
where the leading-order concentration field satisfies the boundary-layer equation \eqref{eq:advection-diffusion_small_y} with boundary conditions
\begin{equation}
\hat{c}_0(0, \hat{y}) = 1, \quad \hat{c}_0(\hat{x}, 0) = 0, \quad
\left. \frac{\partial \hat{c}_0}{\partial \hat{y}} \right|_{\hat{y}=1} = 0. 
\label{eq:bcs_large_Da}
\end{equation}
To reduce the boundary-layer problem to an ODE, we introduce the similarity variable
\begin{equation}
\hat{c}_0 = f(\xi), \quad \xi = \frac{\hat{y}\,\mathrm{Pe}^{1/3}}{\hat{x}^{1/3}}. 
\label{eq:similarity_large_Da}
\end{equation}
Substituting the similarity solution \eqref{eq:similarity_large_Da} into the boundary-layer equation and the boundary conditions \eqref{eq:advection-diffusion_small_y}--\eqref{eq:bcs_large_Da} reduces the problem to the ODE
\begin{equation}
f'' + 2 \xi^2 f' = 0, \quad f(0) = 0, \quad f(\xi \to \infty) \to 1. 
\label{eq:similarity_ode_large_Da}
\end{equation}
The solution to the boundary value problem \eqref{eq:similarity_ode_large_Da} is
\begin{equation}
f = 1 - \frac{\Gamma(1/3, \, 2\xi^3/3)}{\Gamma(1/3)}. 
\label{eq:sol_erf_large_Da}
\end{equation}
Using the asymptotic expansion \eqref{eq:expansion_large_Da} and the leading-order similarity solution \eqref{eq:similarity_large_Da}, the conversion efficiency \eqref{eq:removal_eff} at leading order is
\begin{equation}
\eta = 1 - \int_0^1 \hat{c}_0(\hat{x}=1,\,\hat{y}) \, d\hat{y} + \ldots = \int_0^\infty \frac{\Gamma(1/3, \, 2\xi^3/3)}{\textrm{Pe}^{1/3}\Gamma(1/3)} \, d\xi + \dots.
\label{eq:removal_eff_final}
\end{equation}
Evaluating the integral
\begin{equation}
\int_0^\infty \frac{\Gamma(1/3, \, 2\xi^3/3)}{\textrm{Pe}^{1/3}\Gamma(1/3)} \, d\xi = \frac{(3/2)^{1/3} \Gamma(2/3)}{\textrm{Pe}^{1/3}\Gamma(1/3)},
\label{eq:int_final}
\end{equation}
where the change of variables $\xi = \hat{y} \text{Pe}^{1/3}$ has been used to transform the integral to the similarity variable. 
Substituting the integral result \eqref{eq:int_final} into the conversion efficiency expression \eqref{eq:removal_eff_final} yields the leading-order conversion efficiency
\begin{equation}
\eta = \frac{(3/2)^{1/3} \Gamma(2/3)}{\textrm{Pe}^{1/3}\Gamma(1/3)} + \dots \approx \frac{0.58}{\textrm{Pe}^{1/3}}. 
\label{eq:removal_eff_large_Da_final}
\end{equation}

\section{Methodology for calculating net carbon dioxide equivalent emissions rates from photocatalytic oxidation} \label{sec:Methodology for photocatalyst}

This appendix describes the methodology used to quantify the net climate impact of CH$_4$ PCO on UV-illuminated TiO$_2$. 
The net CO$_2$e emission rate (g/s) associated with PCO is defined as the sum of four terms: (i) direct CO$_2$ produced by the stoichiometric conversion of CH$_4$, (ii) the avoided climate forcing associated with CH$_4$ removal, (iii) indirect CO$_2$ emissions from UV energy use, and (iv) the embodied carbon of the photocatalyst material. 
The parameter values used in these calculations are listed in Table~\ref{tab:parameters_pco}. 
The total net CO$_2$e balance is given by:
\begin{multline} \label{eq:net_co2e_pco}
\text{net CO}_2\text{e emission rate} = \text{CO}_2\text{ production rate} - \text{CO}_2\text{e removal rate} \\ + \text{CO}_2\text{ UV emission rate} + \text{CO}_2\text{e material emission rate},
\end{multline}
with each term defined in Sections~\ref{sec:CO2 produced PCO}--\ref{sec:Embodied carbon PCO}. 

\begin{table}[t!]
\centering
\caption{Parameters used in the life-cycle assessment of TiO$_2$-based PCO of CH$_4$, including catalyst properties, UV energy demand and emissions factors~\cite{kumar2021climate, RioTinto2020Climate, escobedo2020photocatalysis, eea2024electricityco2, jm_scr2020, hammershoi2018impact}. 
}
\small
\renewcommand{\arraystretch}{1.3}
\begin{tabular}{l@{\hskip 40pt}l}
\hline
\textbf{Parameter} & \textbf{Value} \\
\hline
Catalyst material & TiO$_2$ \\
Catalyst area (ventilation panel) & 0.04 m$^2$ \\
UV lamp power & 25 W \\
Operation time & 24 h/day, 365 days/y \\
Electricity CO$_2$ intensity & 200 gCO$_2$/kWh \\
Catalyst lifetime & 1 year \\
TiO$_2$ emission factor & 4 gCO$_2$e/gTiO$_2$ \\
Catalyst mass & 3.2 g \\
\hline
\end{tabular}
\label{tab:parameters_pco}
\end{table}

\subsection{Carbon dioxide production rate}\label{sec:CO2 produced PCO}

The PCO of CH$_4$ follows the reaction \(\text{CH}_4 + 2\text{O}_2 \rightarrow \text{CO}_2 + 2\text{H}_2\text{O}\), which produces one mole of CO$_2$ per mole of CH$_4$ converted. 
The experimentally measured CH$_4$ conversion rate (Section~\ref{sec:Experiments}), reported in molecules per second, is converted to moles per second using Avogadro’s number and multiplied by the molar mass of CO$_2$ (44 g/mol) to obtain the CO$_2$ production rate in g/s. 

\subsection{Carbon dioxide equivalent removal rate}\label{sec:CH4 avoided PCO}

Because CH$_4$ has a high GWP, its conversion results in a net reduction in CO$_2$e. 
Using the 20-year GWP value (GWP = 84~\cite{kumar2021climate}), the avoided emissions per mole of CH$_4$ converted are $(\text{GWP} - 1) \times M_{\rm CO_2}$, where the subtraction of 1 accounts for the single mole of CO$_2$ produced during oxidation. 
Multiplying this factor by the CH$_4$ conversion rate yields the CO$_2$e removal rate in g/s. 

\subsection{Carbon dioxide UV emission rate}\label{sec:UV energy emissions}

Continuous UV illumination of the photocatalyst requires electrical power. 
The electrical energy consumption is calculated as \( E = P_{\text{UV}} \times t_{\text{op}} \), where \(P_{\text{UV}} = 25\) W represents the typical power of UV lamps used in ventilation-integrated PCO systems~\cite{escobedo2020photocatalysis}, excluding high-power lamps ($\ge$500 W) not applicable in this context, and \(t_{\text{op}} = 24 \times 3600 = 86400 \text{ s/day}\). 
The annual electrical demand is converted to CO$_2$ emissions using an electricity emissions factor of 200 gCO$_2$/kWh~\cite{eea2024electricityco2}. 
Dividing the annual CO$_2$ emissions by the total number of seconds in a year yields the CO$_2$ emission rate from UV operation in g/s. 

\subsection{Embodied carbon in photocatalyst materials}\label{sec:Embodied carbon PCO}

The embodied CO$_2$e associated with TiO$_2$ production is accounted for using a material emissions factor of 4 gCO$_2$e/gTiO$_2$ ~\cite{RioTinto2020Climate}. 
The total TiO$_2$ mass is $w=3.2$ g, corresponding to a coating density of 80 g/m$^2$\cite{jm_scr2020} applied over a 0.04 m$^2$ ventilation panel. 
The total embodied emissions, $w e$, are considered as an annual material burden for the system, yielding a CO$_2$e material emission rate in g/s. 

\subsection{Calculation of total net carbon dioxide equivalent emissions}

The net climate impact of the photocatalytic system is calculated by summing all contributions in~\eqref{eq:net_co2e_pco}. 
This expression accounts for the climate benefit of CH$_4$ removal as well as the CO$_2$e contributions from UV operation and catalyst material production. 
Results for the UV intensities, inlet CH$_4$ concentrations, and flow rates investigated experimentally (Section~\ref{sec:Experiments}) are presented in Table~\ref{tab:scaling_up}. 
A negative net CO$_2$e emission rate indicates that the photocatalytic system provides a net climate benefit. 

\printbibliography

@article{zhang2024multivariate,
  title   = {Multivariate modeling of intrinsic kinetics for gas-solid heterogeneous photocatalytic reaction: A general method for different pollutant-photocatalyst systems},
  author  = {Zhang, G. and Liu, J. and Tan, Z. and Yu, H.},
  journal = {Chem. Eng. J.},
  volume  = {479},
  pages   = {147651},
  year    = {2024},
  publisher = {Elsevier}
}

@article{timmerhuis2021connecting,
  title   = {Connecting experimental degradation kinetics to theoretical models for photocatalytic reactors: The influence of mass transport limitations},
  author  = {Timmerhuis, N. A. B. and Wood, J. A. and Lammertink, R. G. H.},
  journal = {Chem. Eng. Sci.},
  volume  = {245},
  pages   = {116835},
  year    = {2021},
  publisher = {Elsevier}
}

@article{malayeri2019modeling,
  title   = {Modeling of volatile organic compounds degradation by photocatalytic oxidation reactor in indoor air: A review},
  author  = {Malayeri, M. and Haghighat, F. and Lee, C.-S.},
  journal = {Build. Environ.},
  volume  = {154},
  pages   = {309--323},
  year    = {2019},
  publisher = {Elsevier}
}

@article{ballari2010modelling,
  title   = {Modelling and experimental study of the NOx photocatalytic degradation employing concrete pavement with titanium dioxide},
  author  = {Ballari, M. M. and Hunger, M. and H{\"u}sken, G. and Brouwers, H. J. H.},
  journal = {Catal. Today},
  volume  = {151},
  number  = {1--2},
  pages   = {71--76},
  year    = {2010},
  publisher = {Elsevier}
}

@article{yu2017experiments,
  title   = {Experiments and kinetics of solar PCO for indoor air purification in PCO/TW system},
  author  = {Yu, B. and He, W. and Li, N. and Zhou, F. and Shen, Z. and Chen, H. and Xu, G.},
  journal = {Build. Environ.},
  volume  = {115},
  pages   = {130--146},
  year    = {2017},
  publisher = {Elsevier}
}

@article{tsopelakou2024exploring,
  title   = {Exploring the bounds of methane catalysis in the context of atmospheric methane removal},
  author  = {Tsopelakou, A. M. and Stallard, J. and Archibald, A. T. and Fitzgerald, S. and Boies, A. M.},
  journal = {Environ. Res. Lett.},
  volume  = {19},
  number  = {5},
  pages   = {054020},
  year    = {2024},
  publisher = {IOP}
}

@misc{jm_scr2020,
  title        = {SCR Catalyst},
  author       = {{Johnson Matthey Catalysts Ltd.}},
  howpublished = {U.S. Patent Application No. US2020/0108376},
  year         = {2020},
  url          = {https://www.freepatentsonline.com/20200108376.pdf}
}

@article{manisalidis2020environmental,
  title   = {Environmental and health impacts of air pollution: A review},
  author  = {Manisalidis, I. and Stavropoulou, E. and Stavropoulos, A. and Bezirtzoglou, E.},
  journal = {Front. Public Health},
  volume  = {8},
  pages   = {14},
  year    = {2020}
}

@book{world2021global,
  title={WHO global air quality guidelines: particulate matter (PM2.5 and PM10), ozone, nitrogen dioxide, sulfur dioxide and carbon monoxide},
  author={{World Health Organization}},
  year={2021},
  publisher={World Health Organization}
}

@article{sosa2017human,
  title   = {Human health risk due to variations in PM10--PM2.5 and associated PAHs levels},
  author  = {Sosa, B. S. and Porta, A. and Lerner, J. E. C. and Noriega, R. B. and Massolo, L.},
  journal = {Atmos. Environ.},
  volume  = {160},
  pages   = {27--35},
  year    = {2017},
  publisher = {Elsevier}
}

@article{kumar2021climate,
  title   = {Climate change and cities: Challenges ahead},
  author  = {Kumar, P.},
  journal = {Front. Sustain. Cities},
  volume  = {3},
  pages   = {645613},
  year    = {2021},
  publisher = {Frontiers}
}

@article{jones2023national,
  title   = {National contributions to climate change due to historical emissions of carbon dioxide, methane, and nitrous oxide since 1850},
  author  = {Jones, M. W. and Peters, G. P. and Gasser, T. and Andrew, R. M. and Schwingshackl, C. and Gutschow, J. and Houghton, R. A. and Friedlingstein, P. and Pongratz, J. and Le Qu{\'e}r{\'e}, C.},
  journal = {Sci. Data},
  volume  = {10},
  number  = {1},
  pages   = {155},
  year    = {2023},
  publisher = {Nature}
}

@article{prentice2001carbon,
  title = {The carbon cycle and atmospheric carbon dioxide},
  author = {Prentice, I. C. and Farquhar, G. D. and Fasham, M. J. R. and Goulden, M. L. and Heimann, M. and Jaramillo, V. J. and Kheshgi, H. S. and Le Quéré, C. and Scholes, R. J. and Wallace, D. W. R. and others},
  journal = {Climate Change 2001: The Scientific Basis, Intergovernmental Panel on Climate Change},
  year = {2001}
}

@article{boningari2016impact,
  title   = {Impact of nitrogen oxides on the environment and human health: Mn-based materials for the NO$_x$ abatement},
  author  = {Boningari, T. and Smirniotis, P. G.},
  journal = {Curr. Opin. Chem. Eng.},
  volume  = {13},
  pages   = {133--141},
  year    = {2016},
  publisher = {Elsevier}
}

@article{zhang2019ozone,
  title   = {Ozone pollution: A major health hazard worldwide},
  author  = {Zhang, J. and Wei, Y. and Fang, Z.},
  journal = {Front. Immunol.},
  volume  = {10},
  pages   = {2518},
  year    = {2019},
  publisher = {Frontiers}
}

@article{randall2024cost,
  title   = {Cost modeling of photocatalytic decomposition of atmospheric methane and nitrous oxide},
  author  = {Randall, R. and Jackson, R. B. and Majumdar, A.},
  journal = {Environ. Res. Lett.},
  volume  = {19},
  number  = {6},
  pages   = {064015},
  year    = {2024},
  publisher = {IOP}
}

@article{tomlinson2025harnessing,
  title   = {Harnessing natural and mechanical airflows for surface-based atmospheric pollutant removal},
  author  = {Tomlinson, S. D. and Tsopelakou, A. M. and Onn, T. M. and Barrett, S. R. H. and Fitzgerald, S. and Boies, A. M.},
  journal = {arXiv},
  pages   = {arXiv:2503.11803},
  year    = {2025}
}

@article{Russell2021Review,
  title   = {A review of photocatalytic materials for urban NOx remediation},
  author  = {Russell, H. S. and Frederickson, L. B. and Hertel, O. and Ellermann, T. and Jensen, S. S.},
  journal = {Catalysts},
  volume  = {11},
  number  = {6},
  pages   = {675},
  year    = {2021},
  publisher = {MDPI}
}

@article{ma2025,
  title   = {Environmental catalytic city: New engine for air pollution control},
  author  = {Ma, J. and Chu, B. and Li, X. and Wang, H. and Ma, Q. and He, G. and Liu, Q. and Wang, S. and He, K. and Zhao, J.},
  journal = {J. Environ. Sci.},
  volume  = {156},
  pages   = {576--583},
  year    = {2025},
  publisher = {Elsevier}
}

@article{escobedo2020photocatalysis,
  title={Photocatalysis for air treatment processes: Current technologies and future applications for the removal of organic pollutants and viruses},
  author={Escobedo, S. and de Lasa, H.},
  journal={Catalysts},
  volume={10},
  number={9},
  pages={966},
  year={2020},
  publisher={MDPI}
}

@misc{eea2024electricityco2,
  author       = {{European Environment Agency}},
  title        = {Greenhouse Gas Emission Intensity of Electricity Generation (g CO$_2$e/kWh)},
  howpublished = {European Environment Agency, European Topic Centre on Climate Change Mitigation and Energy},
  month        = jun,
  year         = {2024},
  note         = {Chart and country-level data, EU average emission intensity},
  url          = {https://www.eea.europa.eu/en/analysis/maps-and-charts/co2-emission-intensity-15}
}

@misc{RioTinto2020Climate,
  author       = {{Rio Tinto}},
  title        = {Our approach to climate change: 2020 Climate Change Report},
  year         = {2020},
  url          = {https://www.riotinto.com/-/media/Content/Documents/Invest/Reports/Climate-Change-reports/RT-climate-scope-123-report.pdf},
  note         = {Accessed: 2025-08-07},
  institution  = {Rio Tinto},
}

@article{hammershoi2018impact,
  title   = {Impact of SO$_2$-poisoning over the lifetime of a Cu-CHA catalyst for NH$_3$-SCR},
  author  = {Hammersh{\o}i, P. S. and Jensen, A. D. and Janssens, T. V. W.},
  journal = {Appl. Catal. B},
  volume  = {238},
  pages   = {104--110},
  year    = {2018},
  publisher = {Elsevier}
}

@article{zhou2025,
  title   = {Effect of sulfur poisoning during worldwide harmonized light vehicles test cycle on NO$_x$ reduction performance and active sites of selective catalytic reduction filter},
  author  = {Zhou, Z. and Yu, F. and Yang, D. and Chang, S. and He, X. and Zhao, Y. and Ma, J. and Chen, T. and Lai, H. and Lin, H.},
  journal = {Catalysts},
  volume  = {15},
  number  = {7},
  pages   = {682},
  year    = {2025},
  publisher = {MDPI}
}

@article{pitchai1986,
  author = {Pitchai, R. and Klier, K.},
  title = {Partial oxidation of methane},
  journal = {Catal. Rev. Sci. Eng.},
  volume = {28},
  number = {1},
  pages = {13--88},
  year = {1986}
}

@article{Dissanayake1991,
  author = {Dissanayake, D. and Rosynek, M. P. and Kharas, K. C. C. and Lunsford, J. H.},
  title = {Partial oxidation of methane to carbon monoxide and hydrogen over a Ni/Al2O3 catalyst},
  journal = {J. Catal.},
  volume = {132},
  number = {1},
  pages = {117--127},
  year = {1991}
}

@article{Anekwe2025,
  author = {Anekwe, I. M. S. and Isa, Y. M.},
  title = {Unlocking catalytic longevity: a critical review of catalyst deactivation pathways and regeneration technologies},
  journal = {Energy Adv.},
  volume = {4},
  number = {9},
  pages = {1075--1113},
  year = {2025}
}

@article{Yoshida2008,
  author = {Yoshida, H. and Hirao, K. and Nishimoto, J.-i. and Shimura, K. and Kato, S. and Itoh, H. and Hattori, T.},
  title = {Hydrogen production from methane and water on platinum loaded titanium oxide photocatalysts},
  journal = {J. Phys. Chem. C},
  volume = {112},
  number = {14},
  pages = {5542--5551},
  year = {2008}
}

@article{Jiang2023,
  author = {Jiang, Y. and Li, S. and Wang, S. and Zhang, Y. and Long, C. and Xie, J. and Fan, X. and Zhao, W. and Xu, P. and Fan, Y.},
  title = {Enabling specific photocatalytic methane oxidation by controlling free radical type},
  journal = {J. Am. Chem. Soc.},
  volume = {145},
  number = {4},
  pages = {2698--2707},
  year = {2023}
}

@article{Luo2021,
  author = {Luo, L. and Gong, Z. and Xu, Y. and Ma, J. and Liu, H. and Xing, J. and Tang, J.},
  title = {Binary Au--Cu reaction sites decorated ZnO for selective methane oxidation to C1 oxygenates with nearly 100\% selectivity at room temperature},
  journal = {J. Am. Chem. Soc.},
  volume = {144},
  number = {2},
  pages = {740--750},
  year = {2021}
}

@article{Song2020,
  author = {Song, H. and Meng, X. and Wang, S. and Zhou, W. and Song, S. and Kako, T. and Ye, J.},
  title = {Selective photo-oxidation of methane to methanol with oxygen over dual-cocatalyst-modified titanium dioxide},
  journal = {ACS Catal.},
  volume = {10},
  number = {23},
  pages = {14318--14326},
  year = {2020}
}

@article{Song2018,
  author = {Song, H. and Meng, X. and Wang, Z.-j. and Wang, Z. and Chen, H. and Weng, Y. and Ichihara, F. and Oshikiri, M. and Kako, T. and Ye, J.},
  title = {Visible-light-mediated methane activation for steam methane reforming under mild conditions: a case study of Rh/TiO$_2$ catalysts},
  journal = {ACS Catal.},
  volume = {8},
  number = {8},
  pages = {7556--7565},
  year = {2018}
}

@article{Wei2004a,
  author = {Wei, J. and Iglesia, E.},
  title = {Reaction pathways and site requirements for the activation and chemical conversion of methane on Ru-based catalysts},
  journal = {J. Phys. Chem. B},
  volume = {108},
  number = {22},
  pages = {7253--7262},
  year = {2004}
}

@article{Wei2004b,
  author = {Wei, J. and Iglesia, E.},
  title = {Mechanism and site requirements for activation and chemical conversion of methane on supported Pt clusters and turnover rate comparisons among noble metals},
  journal = {J. Phys. Chem. B},
  volume = {108},
  number = {13},
  pages = {4094--4103},
  year = {2004}
}

@article{Hu2023,
  author = {Hu, D. and Addad, A. and Tayeb, K. B. and Ordomsky, V. V. and Khodakov, A. Y.},
  title = {Thermocatalysis enables photocatalytic oxidation of methane to formic acid at room temperature beyond the selectivity limits},
  journal = {Cell Rep. Phys. Sci.},
  volume = {4},
  number = {2},
  year = {2023}
}

@article{Fan2021,
  author = {Fan, Y. and Zhou, W. and Qiu, X. and Li, H. and Jiang, Y. and Sun, Z. and Han, D. and Niu, L. and Tang, Z.},
  title = {Selective photocatalytic oxidation of methane by quantum-sized bismuth vanadate},
  journal = {Nat. Sustain.},
  volume = {4},
  number = {6},
  pages = {509--515},
  year = {2021}
}

@article{Li2015,
  author = {Li, H. and Lei, Z. and Liu, C. and Zhang, Z. and Lu, B.},
  title = {Photocatalytic degradation of lignin on synthesized Ag--AgCl/ZnO nanorods under solar light and preliminary trials for methane fermentation},
  journal = {Bioresour. Technol.},
  volume = {175},
  pages = {494--501},
  year = {2015}
}

@article{Li2011,
  author = {Li, L. and Li, G.-D. and Yan, C. and Mu, X.-Y. and Pan, X.-L. and Zou, X.-X. and Wang, K.-X. and Chen, J.-S.},
  title = {Efficient Sunlight-Driven Dehydrogenative Coupling of Methane to Ethane over a Zn+-Modified Zeolite},
  journal = {Angew. Chem.},
  volume = {123},
  number = {36},
  year = {2011}
}

@article{Han2019,
  author = {Han, B. and Wei, W. and Li, M. and Sun, K. and Hu, Y. H.},
  title = {A thermo-photo hybrid process for steam reforming of methane: highly efficient visible light photocatalysis},
  journal = {Chem. Commun.},
  volume = {55},
  number = {54},
  pages = {7816--7819},
  year = {2019}
}

@article{Pennacchio2024,
  author = {Pennacchio, L. and Mikkelsen, M. K. and Krogsb{\o}ll, M. and van Herpen, M. and Johnson, M. S.},
  title = {Physical and practical constraints on atmospheric methane removal technologies},
  journal = {Environ. Res. Lett.},
  volume = {19},
  number = {10},
  pages = {104058},
  year = {2024}
}

@article{Tomlinson2026,
  author = {Tomlinson, S. D. and Tsopelakou, A. M. and Onn, T. M. and Barrett, S. R. H. and Boies, A. M. and Fitzgerald, S.},
  title = {Modelling laminar flow in V-shaped filters integrated with catalyst technologies for atmospheric pollutant removal},
  journal = {Int. J. Heat Mass Transfer},
  volume = {257},
  pages = {128206},
  year = {2026}
}

@article{krogsboll2025efficient,
  author = {Krogsb{\o}ll, M. and Rezaei, M. and Fogde, N. and Weiss, N. D. and Russell, H. S. and Feilberg, A. and Johnson, M. S.},
  title = {Efficient mitigation of dilute methane, ammonia, and odor in ventilation air from cow and pig barns and a biogas plant: photoreactor field demonstration},
  journal = {ACS ES\&T Air},
  volume = {2},
  number = {8},
  pages = {1648--1655},
  year = {2025},
  publisher = {ACS Publ.}
}

@article{iversen2025mitigation,
  title={Mitigation of atmospheric and elevated methane by photochemical oxidation at ambient conditions},
  author={Iversen, N. and Roslev, P.},
  journal={Sci. Total Environ.},
  volume={976},
  pages={179338},
  year={2025},
  publisher={Elsevier}
}

@article{abernethy2023assessing,
  author = {Abernethy, S. and Kessler, M. I. and Jackson, R. B.},
  title = {Assessing the potential benefits of methane oxidation technologies using a concentration-based framework},
  journal = {Environ. Res. Lett.},
  volume = {18},
  number = {9},
  pages = {094064},
  year = {2023},
  publisher = {IOP Publ.}
}

@article{hickey2024economics,
  title = {Economics of Enhanced Methane Oxidation Relative to Carbon Dioxide Removal},
  author = {Hickey, C. and Allen, M.},
  journal = {Environ. Res. Lett.},
  volume = {19},
  number = {6},
  pages = {064043},
  year = {2024},
  publisher = {IOP Publishing}
}

@article{rauchenwald2020new,
  title={New method of destroying waste anesthetic gases using gas-phase photochemistry},
  author={Rauchenwald, V. and Rollins, M. D. and Ryan, S. M. and Voronov, A. and Feiner, J. R. and {\v{S}}arka, K. and Johnson, M. S.},
  journal={Anesth. Analg.},
  volume={131},
  number={1},
  pages={288--297},
  year={2020},
  publisher={LWW}
}

@article{kessler2026humidity,
  title={A Humidity-Tolerant Photocatalyst for Methane Removal},
  author={Kessler, M. I. and Randall, R. and Wan, G. and Xu, K. and Zhang, Y. and Dionne, J. A. and Jackson, R. B. and Majumdar, A.},
  journal={Environ. Sci. Technol.},
  volume={60},
  number={7},
  pages={5442--5452},
  year={2026},
  publisher={ACS Publications}
}

@article{marina2026evaluating,
  title={Evaluating the potential of thermal catalysis for environmental methane mitigation},
  author={Tsopelakou, A. M. and Tomlinson, S. D. and Fitzgerald, S. and Onn, T. M and Boies, A. M.},
  journal={Sustain. Sci. Technol.},
  volume={3},
  number={2},
  pages={024002},
  year={2026},
  publisher={IOP Publishing}
}

@manual{matlab2025,
  title        = {MATLAB},
  organization = {The MathWorks, Inc.},
  address      = {Natick, Massachusetts},
  year         = {2025}
}

@article{ying2025unraveling,
  title={Unraveling humidity's dual effects on trace methane photocatalytic ozonation: Enhanced radical generation versus active site blockade},
  author={Ying, J. and Wang, Y. and Zhang, H. and Bai, Y. and Zhu, J. and Lu, X. and Li, W. and Mu, L.},
  journal={J. Environ. Chem. Eng.},
  pages={120330},
  year={2025},
  publisher={Elsevier}
}

@article{argyle2015heterogeneous,
  title={Heterogeneous catalyst deactivation and regeneration: a review},
  author={Argyle, M. D. and Bartholomew, C. H.},
  journal={Catalysts},
  volume={5},
  number={1},
  pages={145--269},
  year={2015},
  publisher={MDPI}
}

@article{thornton2022impact,
  author = {Thornton, G. M. and Fleck, B. A. and Fleck, N. and Kroeker, E. and Dandnayak, D. and Zhong, L. and Hartling, L.},
  title = {The impact of heating, ventilation, and air conditioning design features on the transmission of viruses, including the 2019 novel coronavirus: A systematic review of ultraviolet radiation},
  journal = {PLoS One},
  volume = {17},
  number = {4},
  pages = {e0266487},
  year = {2022},
  publisher = {Public Library of Science San Francisco, CA USA}
}

@techreport{eia2022cbecshighlights,
  author       = {{U.S. Energy Information Administration}},
  title        = {2018 {Commercial Buildings Energy Consumption Survey (CBECS)}: Consumption and Expenditures Highlights},
  institution  = {U.S. Department of Energy, Office of Energy Statistics},
  year         = {2022},
  month        = dec,
  howpublished = {\url{https://www.eia.gov/consumption/commercial/}},
}

@misc{frank2017incropera,
  title={Incropera’s Principles of Heat and Mass Transfer},
  author={Incropera, F. P.},
  year={2017},
  publisher={Wiley: Hoboken, NJ, USA}
}

\end{document}